\documentclass[fleqn,usenatbib]{mnras}
\usepackage{graphicx}	
\usepackage{amsmath}	
\usepackage[dvipsnames]{xcolor}  
\usepackage{amssymb}	
\usepackage{macros_vdb} 
\begin{document}
%

\title[Tidal Evolution of Substructure]
      {Dark Matter Substructure in Numerical Simulations: \\ $\;\;\;$A Tale of
        Discreteness Noise, Runaway Instabilities, \\
        $\;\;\;$and Artificial Disruption}

\author[van den Bosch et al.]{%
   Frank~C.~van den Bosch$^1$\thanks{E-mail: frank.vandenbosch@yale.edu},
   Go Ogiya$^2$, 
\vspace*{8pt}
\\
   $^1$Department of Astronomy, Yale University, PO. Box 208101, New Haven, CT 06520-8101\\
   $^2$Laboratoire Lagrange, Observatoire de la C\^ote d'Azur, CNRS, Blvd de l'Observatoire, CS 34229, F-06304  Nice cedex 4, France}


\date{}

\pagerange{\pageref{firstpage}--\pageref{lastpage}}
\pubyear{2013}

\maketitle

\label{firstpage}


\begin{abstract}
  To gain understanding of the complicated, non-linear and numerical 
  processes associated with the tidal evolution of dark matter subhaloes 
  in numerical simulation, we perform a large suite of idealized simulations 
  that follow individual $N$-body subhaloes in a fixed, analytical host 
  halo potential. By varying both physical and numerical parameters, we 
  investigate under what conditions the subhaloes undergo disruption. We 
  confirm the conclusions from our more analytical assessment in van den 
  Bosch et al. that most disruption is numerical in origin; as long 
  as a subhalo is resolved with sufficient mass and force resolution, a 
  bound remnant survives. This implies that state-of-the-art cosmological 
  simulations still suffer from significant overmerging. We demonstrate 
  that this is mainly due to inadequate force-softening, which causes 
  excessive mass loss and artificial tidal disruption. In addition, we 
  show that subhaloes in $N$-body simulations are susceptible to a runaway 
  instability triggered by the amplification of discreteness noise in the 
  presence of a tidal field. These two processes conspire to put serious 
  limitations on the reliability of dark matter substructure in 
  state-of-the-art cosmological simulations. We present two criteria that 
  can be used to assess whether individual subhaloes in cosmological 
  simulations are reliable or not, and advocate that subhaloes that satisfy 
  either of these two criteria be discarded from further analysis. We 
  discuss the potential implications of this work for several areas in 
  astrophysics.
\end{abstract} 


\begin{keywords}
instabilities ---
methods: numerical ---
galaxies: haloes --- 
cosmology: dark matter
\end{keywords}


\section{Introduction} 
\label{sec:intro}

Prior to 1997, numerical $N$-body simulations of structure formation
in a cold dark matter (CDM) cosmology suffered from a serious
`overmerging' problem in that simulated CDM haloes revealed little to
no substructure, in clear contrast with the wealth of `substructure'
(i.e., satellite galaxies) observed in galaxy groups and clusters.
While some speculated that baryonic physics would resolve this problem
\citep[e.g.,][]{Frenk.etal.88}, others argued that it is a numerical
artifact arising from insufficient mass and/or force resolution
\citep[e.g.,][]{Carlberg.94, vKampen.95, vKampen.00a, Moore.etal.96a,
  Klypin.etal.99a}. Indeed, with increasing resolution the simulations
started to reveal a wealth of substructure
\citep[e.g.,][]{Tormen.etal.97, Brainerd.etal.98, Moore.etal.98b}, and
by the close of the last millennium the `overmerging problem' had been
superseded by the `missing satellite problem' \citep{Klypin.etal.99b,
  Moore.etal.99}. Over the years it has become clear that the latter
is mainly a manifestation of poorly understood baryonic physics
related to galaxy formation \citep[see][for a comprehensive
  review]{Bullock.BoylanKolchin.17}, and the astrophysical community
has become more and more confident that the $N$-body simulations make
reliable predictions regarding the abundance and demographics of dark
matter substructure \citep[e.g.,][]{Ghigna.etal.98,  
Klypin.etal.99a, Diemand.etal.04, Gao.etal.04, Kravtsov.etal.04, 
Giocoli.etal.08, Giocoli.etal.10}. In particular, a number of resolution 
and comparison studies  \citep[e.g.,][]{Springel.etal.08, Onions.etal.12,
Knebe.etal.13, vdBosch.Jiang.16, Griffen.etal.16} have shown that subhalo 
mass functions  are converged\footnote{A common practice with numerical 
simulations is to perform `convergence studies' in which simulations are 
ran at different resolutions. Those results that are robust to an increase 
in resolution are deemed `converged'.} down to 50-100 particles per subhalo.

Does this mean that, in modern simulations, numerical overmerging only
occurs when the subhalo has less than $\sim 50$ particles?  Subhalo
disruption is still extremely prevalent in modern simulations, with
inferred fractional disruption rates (at $z=0$) of $\sim 13$ percent
per Gyr \citep[][]{Diemand.Moore.Stadel.04, vdBosch.17}. This implies
that $\sim 65$ ($90$) percent of all subhaloes accreted around $z=1$
($z=2$) are disrupted by $z=0$ \citep[][]{Han.etal.16,
  Jiang.vdBosch.17}. As discussed in \citet{vdBosch.17}, roughly 20
percent of this disruption occurs above the mass resolution limit of
50 particles; in fact, the mass function of disrupting subhaloes is
indistinguishable from that of the surviving population.

What is the dominant cause of this prevalent disruption of subhaloes
in numerical simulations? In particular, is it artificial (numerical)
or real (physical)?  Based on the fact that simulations seem to yield
consistent, converged results for the mass function (and spatial
distribution) of subhaloes above a resolution limit of 50-100
particles, one is tempted to conclude that any disruption of subhaloes
above this `resolution limit' must be physical in origin
\citep[see][for a detailed discussion]{Diemand.Moore.Stadel.04}.
However, convergence is only a necessary, but not a sufficient
condition to guarantee that the results are reliable. In addition,
there is no consensus as to what physical mechanism dominates, with
most studies arguing either for tidal heating or tidal stripping. We
have therefore initiated a comprehensive study aimed at answering
these questions. In \citet[hereafter Paper I]{vdBosch.etal.17} we use
both analytical estimates and idealized numerical simulations to
investigate whether subhalo disruption is mainly physical, due to
tidal heating and stripping, or numerical (i.e., artificial). We show
that, in the absence of baryonic processes, the complete, physical
disruption of CDM substructure is extremely rare, and that most
disruption in numerical simulations therefore must be artificial. We
discuss various processes that have been associated with numerical
overmerging, and conclude that inadequate force-softening is the most
likely culprit.

In this paper, we use a large suite of idealized numerical simulations
and experiments to examine the tidal evolution and disruption of
subhaloes in unprecedented detail. We confirm the conclusions from
Paper~I and demonstrate that state-of-the-art numerical $N$-body
simulations indeed suffer from significant overmerging, mainly driven
by inadequate force-softening. In addition, we show that subhaloes in
$N$-body simulations are susceptible to a runaway instability which is
triggered by the amplification of discreteness noise in the presence
of a tidal field.

The main goal of this paper is to address the following questions:
\begin{itemize}
\item under what conditions does numerical disruption occur, and what
  causes it?
\item what are the numerical requirements (i.e., number of particles,
  softening length, time stepping, etc.) to properly resolve the tidal
  evolution of dark matter substructure.
\end{itemize}
Addressing these questions is facilitated by considering simplified
settings, and we therefore resort to using idealized numerical
simulations, in which we represent the subhalo by a $N$-body system,
which we integrate in a static, analytical, external potential
representing the host halo.  We further simplify matters by 
predominantly considering circular orbits, and by focusing exclusively 
on dark matter (i.e., we ignore the potential impact of baryons). The goal of
these idealized, numerical experiments is not to simulate realistic
astrophysical systems, but rather to gain a physical understanding of
the complicated, non-linear and numerical processes associated with
the tidal stripping of dark matter subhaloes.  Although the fraction
of subhaloes/satellite galaxies on purely circular orbits is
vanishingly small \citep[e.g.][]{Zentner.etal.05, Khochfar.Burkert.06, 
Wetzel.11, vdBosch.17}, and realistic host haloes are not static in that
they respond dynamically to the presence of the subhalo, these
idealizations have the advantage that they minimize the impact of
tidal heating and dynamical friction, thereby allowing us to focus on
the impact of tidal stripping. Most of the simulations presented below
have been run for more than 60 Gyr, much longer than the age of the
Universe. This stresses once more that we are striving to gain
physical understanding, not necessarily to describe or model a
realistic setting.

This paper is organized as follows. In \S\ref{sec:method} we describe
the simulations, the initial conditions, and the treecodes and method
used to run and analyze the simulations. \S\ref{sec:stripping}
presents a case study of one of our idealized simulations,
highlighting the typical tidal evolution of a dark matter subhalo. In
\S\ref{sec:num} we assess the impact of four numerical parameters that
control the accuracy of the simulations: the time step used to
integrate the equations of motion, the tree opening angle used in the
force calculation, the force softening length, and the actual number
of particles used to simulate the subhalo in question.
\S\ref{sec:convergence} presents a detailed convergence study,
highlighting the impact of insufficient force softening and a
discreteness-noise driven runaway instability. In \S\ref{sec:criteria}
we examine under what conditions individual subhaloes in
state-of-the-art cosmological simulations may be deemed `resolved', 
\S\ref{sec:ICsens} discusses the sensitivity of our results to various 
aspects of our (oversimplified) initial conditions, and \S\ref{sec:caveats} 
highlights some potential caveats of our study. Finally, \S\ref{sec:concl} 
summarizes our results.

Throughout we adopt a Hubble parameter $H_0 = 70 \kmsmpc$, which
corresponds to a Hubble time of $t_\rmH = H^{-1}_0 = 13.97\Gyr$.


\section{Methodology}
\label{sec:method}

\subsection{Initial Conditions}
\label{sec:ICs}

In this paper we simulate individual dark matter haloes (hereafter the
`subhalo' or `satellite') orbiting in a fixed, external potential
(hereafter the `host' halo). Our goal is to investigate how much mass
is stripped from the subhaloes and under what conditions the satellite
disrupts (i.e., has no bound structure surviving).

Both the host halo and the {\it initial} (prior to the onset of tidal
stripping) subhalo are assumed to be spherical, and to have a NFW
density profile
\begin{equation}
  \rho(r) = \rho_\rms \, \left(\frac{r}{r_\rms}\right)^{-1} \,
  \left(1 + \frac{r}{r_\rms}\right)^{-2}\,,  
\end{equation}
\citep[][]{Navarro.etal.97}, where $r_\rms$ is the characteristic
scale radius. We define the virial radius, $\rvir$, as the radius
inside of which the average density is $\Delta_{\rm vir} = 97$ times
the critical density for closure \citep{Bryan.Norman.98}, and the halo
concentration $c \equiv \rvir/r_\rms$. The virial mass of a halo,
$\Mvir$, is defined as the mass inside $\rvir$, while the virial
velocity is defined as the circular velocity at the virial radius,
$V_{\rm vir} = \sqrt{G \Mvir/\rvir}$. The crossing time for such a
halo is
\begin{equation}\label{tcross}
t_{\rm cross} \equiv \frac{\rvir}{\Vvir} = 2.006 \Gyr\,.
\end{equation}

We generate initial conditions (ICs) assuming that the NFW subhalo has
an isotropic velocity distribution, such that its distribution
function (DF) depends only on energy, i.e., $f = f(E)$.  We use the
method of \cite{Widrow.00} to sample particles from the DF using the
standard acceptance-rejection technique \citep[][]{Press.etal.92,
  Kuijken.Dubinski.94}. We truncate the initial subhalo at a
radius $\rmax$, which is a free parameter of our model. Unless
specifically stated otherwise, we adopt $\rmax = \rvir$.
Unfortunately, the DF that we use to generate the ICs is computed
using the \cite{Eddington.16} inversion equation, assuming that the
halo extends to infinity.  Consequently, unless $\rmax = \infty$ the
initial system is not going to be in perfect equilibrium (see App.~B
of Paper I).  \cite{Kazantzidis.etal.04} have suggested a way around
this problem; following \cite{Springel.White.99}, they introduce an
exponential cut-off for $r>r_{\rm max}$, and compute $f(E)$ from this
modified density distribution using the Eddington equation.  The 
cut-off sets in at $\rmax$ and exponentially decreases the density over 
a scale of $r_{\rm decay}$, which is a free parameter that controls the 
sharpness of the transition.  However, since we will embed our haloes in an
external tidal field, with a corresponding tidal radius that lies well
inside of $\rmax$, there is little virtue to such an exponential
cut-off, and to having a halo whose outskirts are in perfect
equilibrium; even if the system were to be in perfect equilibrium in
isolation, the moment we instantaneously introduce it to its tidal
environment, it will no longer be in equilibrium.

\cite{Choi.etal.07, Choi.etal.09} try to account for the tidal field
in their initial conditions, by truncating the halo at the radius
beyond which subhalo particles on a circular orbit become unbound, and
by using Eddington inversion to compute the corresponding DF. However,
their method is also an approximation at best, as their simply cannot 
be a spherical, isotropic equilibrium solution for a system inside a
tidal field.

The experiments conducted here correspond to idealized set-ups that
one will never encounter in nature. In reality, subhaloes will already
have been affected by the tidal field of the host halo well before it
reaches the starting point of our simulation. If realism is the goal
of the simulation, one has little choice but to simulate the system in
its proper cosmological setting (i.e., run a cosmological simulation).
The goal of the idealized experiments described here, though, is to
gain a physical understanding of the complicated, non-linear and
numerical processes associated with the tidal stripping of dark matter
subhaloes. Throughout the paper, we will comment on where the
idealizations and approximations made may potentially impact the
results and conclusions of our study. In particular,
\S\ref{sec:ICsens} presents a detailed discussion of how our ICs
affect the outcome of the simulations.

\subsection{Numerical Simulations}
\label{sec:numsim}

All simulations described in this paper have been carried out using
one of two different $N$-body tree codes. The first is a modified
version of the hierarchical $N$-body code \treecode, written by Joshua
Barnes with some improvements due to John Dubinski. \treecode uses a
\cite{Barnes.Hut.86} octree to compute accelerations based on a
multi-pole expansion up to quadrupole order, and uses a straightforward
second order leap-frog integration scheme to solve 
the equations of motion. Forces between particles are softened using 
a simple Plummer softening. The second code is specifically 
designed for graphic processing unit (GPU) clusters 
\citep{Ogiya.etal.13}. Following the {\tt OTOO} code developed by
\citet{Nakasato.etal.12}, CPU cores construct oct-tree structures of 
$N$-body particles, while GPU cards compute gravitational accelerations 
through tree traversal. \cite{Ogiya.etal.13} improved the algorithm of 
tree traversal proposed by \cite{Nakasato.etal.12}, increasing
the speed of the GPU computations by a factor of 4. Hereafter, we 
refer this code as \GPUtree, which differs from \treecode in that it 
(i) uses a second-order Runge-Kutta integrator and (ii) computes 
accelerations by defining the position of the cell as its centre of mass. 
Hence they are accurate up to dipole order.  As we demonstrate in 
\S\ref{sec:stripping}, despite these differences both codes yield results 
that are in excellent agreement; generally, we use the faster \GPUtree 
whenever we use simulations with $\Np > 10^6$, and \treecode otherwise.

Throughout we adopt model units in which the gravitational constant,
$G$, the initial scale radius, $\rs$, and the initial virial mass of
the subhalo, $\ms$, are all unity. With this choice, the initial
virial velocity and crossing time of the subhalo are $\Vvir =
1/\sqrt{c_\rms}$ and $t_{\rm cross} = c_\rms^{3/2}$, respectively,
with $c_\rms$ the subhalo's NFW concentration parameter. Unless stated
otherwise, we restrict ourselves to subhaloes with $c_\rms=10$, for
which $t_{\rm cross} = 31.6$. Based on Eq.~(\ref{tcross}) we thus have
that a time interval of $\Delta t = 1$ (model units) corresponds to
$63.4 \Myr$. Throughout we adopt an NFW host halo of mass $M_\rmh =
1000 \ms$.  In the $\Lambda$CDM cosmology, to good approximation,
concentration scales with halo mass as $c \propto M^{0.1}$
\citep[e.g.,][]{Dutton.Maccio.14}. Hence, for a mass ratio of 1000,
the ratio in concentration parameters is roughly 2, and we therefore
adopt a concentration for the host halo of $c_\rmh = 5$. Note that the
host halo is always modeled as a fixed, external potential, while the
subhalo is a live $N$-body system.

Unless specifically stated otherwise, we adopt a softening length
given by
\begin{equation}\label{softlength}
\varepsilon = 0.05 \, \left(\frac{\Np}{10^5}\right)^{-1/3}\,.
\end{equation}
As discussed in \S\ref{sec:softening}, this is the optimal softening
length for an NFW halo with $\Np=10^5$ and $c_\rms = 10$ in isolation,
while the scaling with $\Np$ is motivated by \cite{vKampen.00a}. A
detailed discussion as to how the choice of $\varepsilon$ impacts the
simulations results is presented in \S\ref{sec:softening}.  All
simulations are run with a fixed time step of $\Delta t = 0.02$.  As
discussed in \S\ref{sec:timestep} this is an extremely conservative
choice, which guarantees that we resolve the orbital time at the
radius equal to the softening length with more than 30 steps. Finally,
we adopt an opening angle for the force calculations equal to
$\theta=0.7$, which, as we demonstrate in \S\ref{sec:theta} is also
conservative, in that our results are extremely robust to changes in
this parameter. For ease of reference, Table~1 summarizes the physical 
and numerical parameters of our `fiducial' set of simulations.
\begin{table}\label{tab:Fiducial}
\caption{Fiducial Parameters}
\begin{center}
\begin{tabular}{lcc}
\hline\hline
description                    & symbol        & value \\ 
\hline
{\bf Physical Parameters}      &               & \\
\hline
initial mass of subhalo        & $\ms$        & $1$ \\
mass of host halo              & $M_\rmh$       & $1000$ \\
concentration of subhalo       & $c_\rms$       & $10$ \\
concentration of host halo     & $c_\rmh$       & $5$ \\
\hline\hline
{\bf Numerical Parameters}     &               & \\
\hline
number of simulation particles & $\Np$         & $10^5$ \\
simulation time step           & $\Delta t$    & $0.02$ \\
softening length (Plummer)     & $\varepsilon$ & $0.05$ \\
tree opening angle             & $\theta$      & $0.7$ \\
\hline\hline
\end{tabular}
\end{center}
\medskip
\begin{minipage}{\hssize}
  Parameters of our fiducial set of simulations expressed in model
  units for which $G = \ms = \rs = 1$.
\end{minipage}
\end{table}

\subsection{Analysis}
\label{sec:analysis}

One of the main goals of this paper is to study how the bound mass of
subhaloes evolve with time. We define
\begin{equation}
  \fbound(t) \equiv \frac{m_\rms(t)}{\ms} =
  \frac{N_{\rm bound}}{\Np} 
\end{equation}
Here $m_\rms(t)$ is the bound mass of the subhalo at time $t$,
$\ms(r)$ is the {\it initial} (prior to being exposed to the tidal
field of the host halo) mass of the subhalo inside radius $r$, $N_{\rm
  bound}$ is the number of bound particles at time $t$, and $\Np$ is
the total number of particles used in the simulation. 

Determining $\fbound(t)$ is non-trivial. We consider a particle
$i$ to be bound to the subhalo if it's binding energy
\begin{equation}\label{Ebound}
E_i \equiv \frac{1}{2} m_i v^2_{{\rm int},i} - \sum_{j\ne i}
\frac{G \, m_i \, m_j \, \calQ_j}{(\vert \br_j - \br_i \vert^2 + \varepsilon^2)^{1/2}} < 0\,.
\end{equation}
Here $\calQ_j$ is equal to 1 (0) if particle $j$ is bound
(unbound). Clearly, in order to be able to compute $E_i$, one first
needs to know which are the bound particles. This can only be solved
using an iterative scheme. Unfortunately, there is no unique way of
performing this `unbinding' operation, and this is where different
analyses of the same simulation can cause significant differences
\citep[e.g.,][]{Muldrew.etal.11, Knebe.etal.11, Han.etal.12,
  Han.etal.17}. In our analysis we proceed as follows:
\begin{enumerate}
\item We begin by making an initial guess for the boundness,
  $\calQ_i$, of each particle.  At $t=0$ (ICs) we assume that each
  particle is bound ($\calQ_i = 1 \,\,\, \forall i=1,...,\Np$), while at
  later times we assume that $\calQ_i$ is the same as in the previous
  simulation output.
\item For each particle we compute $E_i$ using
  Eq.~(\ref{Ebound}).  We update $\calQ_i$ accordingly and compute the
  new, bound fraction $\fbound = \frac{1}{\Np} \, \sum_i \calQ_i$.
\item Compute the centre-of-mass position and velocity of the halo,
  $\br_{\rm com}$ and $\bv_{\rm com}$, as the average position and
  velocity of the $N = {\rm MAX}(N_{\rm min},f_{\rm com} \, N_{\rm bound})$ 
  particles that are most bound. Here $N_{\rm min}$ and $f_{\rm com}$ 
  are free parameters.
\item Update $\bv_i$ for each particle using the new
  centre-of-mass velocity.
\item Go back to (ii) and iterate until the changes in $\br_{\rm com}$
  and $\bv_{\rm com}$ are smaller than $10^{-4} \rvir$ and $10^{-4}
  \Vvir$, respectively. This typically requires 3-10 iterations.
\end{enumerate}
When computing the gravitational potential term of Eq.~(\ref{Ebound})
we use a Barnes \& Hut octree with the same opening angle, $\theta$,
as in the simulation. We also adopt the same softening. Generally,
smaller values of $f_{\rm com}$ results in a more noisy time-evolution
of $\br_{\rm com}$ and $\bv_{\rm com}$, while for larger values of
$f_{\rm com}$ it is more difficult to trace $\fbound(t)$ when it
becomes small. After careful testing, we obtain stable results for
$N_{\rm min} = 10$ (we never use fewer than 10 particles to determine
the centre-of-mass properties\footnote{Except when $N_{\rm bound}<10$,
  in which case we adopt $N=N_{\rm bound}$.}) and $f_{\rm com} = 0.05$
(centre-of-mass properties are determined using the 5 percent most
bound particles), which are the parameters we use throughout.


\section{Tidal Stripping on Circular Orbits}
\label{sec:stripping}

We start by considering the tidal evolution of NFW subhaloes on
circular orbits in the (static) tidal field of an external host halo.
The choice for circular orbits is motivated by the fact that in this
case the tidal field strength is constant, thus minimizing the
contribution from tidal shocking. As discussed in Paper~I, this is,
somewhat surprisingly, the most difficult case to treat analytically.
Briefly, the main reason is that once some matter is stripped, the
remaining remnant is no longer in virial equilibrium. It responds by
re-virializing, which causes it to expand. This expansion, and the
corresponding decrease in the tidal radius, causes additional matter
to be stripped, bringing the system once again out of virial
equilibrium. In a fixed tidal field (such as on the circular orbits
considered here), this sequence of stripping and re-virialization
basically continues {\it ad infinitum}. Since there is no robust, analytical
theory for how to treat (re-)virialization, i.e., to compute how the
density distribution of the subhalo responds to its outer layers being
stripped\footnote{Several authors have attempted approximate
  treatments \citep[e.g.,][]{Taylor.Babul.01, Penarrubia.Benson.05,
    Pullen.etal.14}, but these approaches are crude at best, and still
  require calibration and validation based on numerical simulations.}, 
it is not possible to develop a completely analytical treatment of tidal
stripping. This is further exacerbated by the ill-defined nature of
the tidal radius (see Paper~I for a detailed discussion) and the
self-friction briefly discussed in \S\ref{sec:dynfric} below. As a
consequence, we cannot validate the results from numerical simulations
against an analytical test-case, and we are instead forced to validate
the simulation results against each other, which is one of the main
focuses of this paper.
\begin{figure*}
\includegraphics[width=\hdsize]{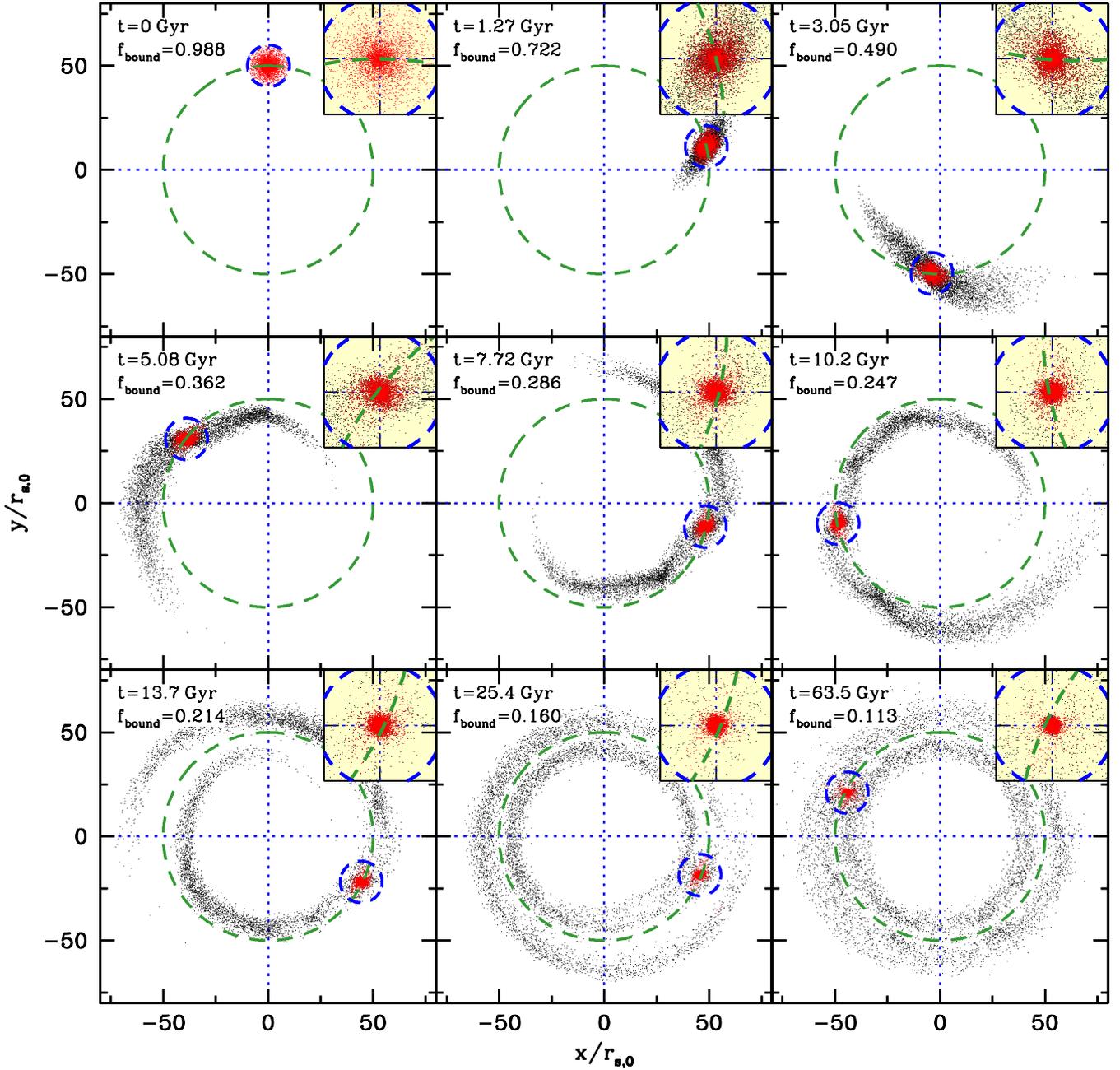}
\caption{Snapshots of random subsets of subhalo particles in one of
  our fiducial simulations (see Table~1 for parameters), projected
  onto the orbital plane of the subhalo. Bound and unbound particles
  are indicated by red and black dots, respectively.  The initial
  subhalo is truncated at its own virial radius, indicated by the
  blue, dashed circle, and reproduced in each panel for reference. The
  subhalo is placed on a circular orbit of radius $\rorbNorm = 0.5$,
  indicated by the green, dashed circle. Each panel indicates the time
  (in Gyr) and the bound fraction, $\fbound$, while the inset shows a
  zoom-in on the central region of the subhalo.}
\label{fig:example}
\end{figure*}

Unless specifically stated otherwise, all subhaloes are initially
truncated at their virial radius, and instantaneously introduced to
the tidal field of the host halo (see \S\ref{sec:ICsens} for
discussion on the sensitivity to these initial conditions).
We put the subhalo on a circular orbit of radius $\rorb$, and
integrate the system for $50,000$ timesteps of $\Delta t = 0.02$ each,
corresponding to a total integration time of $\Delta t = 1000 = 
63.4 \Gyr$ (roughly 4.5 times the age of the Universe). Note that 
$\rorb$ is defined as the distance between the centres-of-mass of 
host and subhalo.
\begin{figure*}
\includegraphics[width=\hdsize]{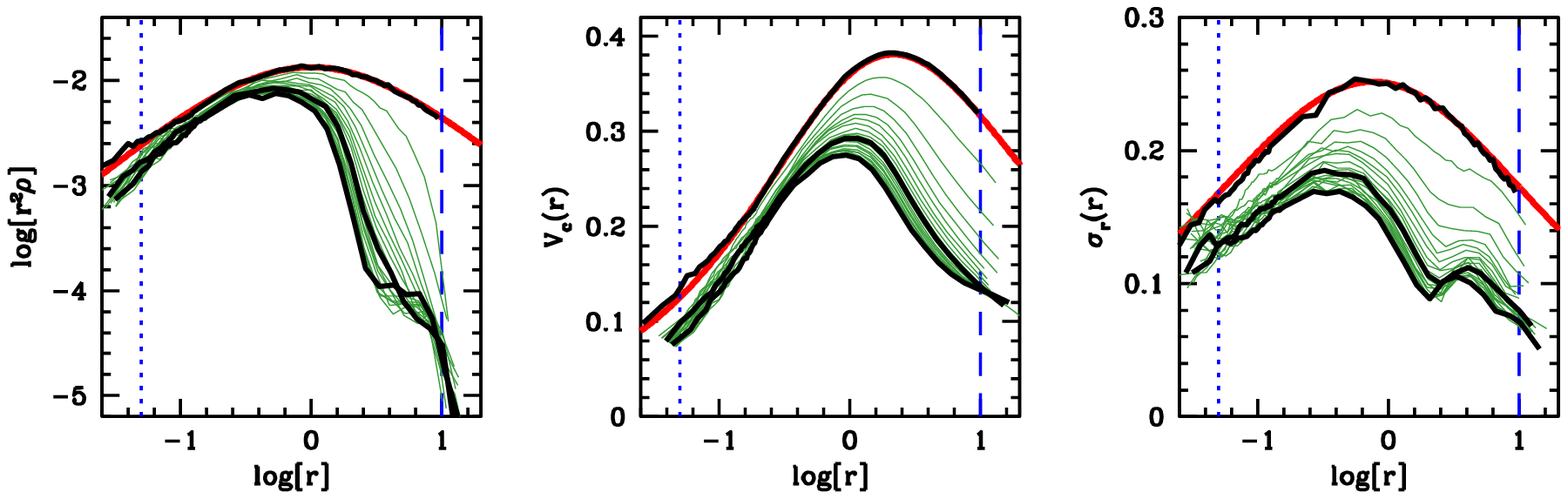}
\caption{Density (left-hand panel), circular velocity (middle panel),
  and radial velocity dispersion (right-hand panel) profiles (in model
  units) of the bound subhalo particles in the same simulation as
  shown in Fig.~\ref{fig:example}. In the left-hand we plot 
  $\log[r^2\rho(r)]$, rather than $\log[\rho(r)]$ to better 
  highlight small differences. Thin and thick lines show the
  profiles ever 2.5 and 25Gyr, respectively, while the thick red
  lines indicate the analytical profiles. The vertical dotted and
  dashed lines mark the softening length and the initial virial
  radius, respectively.}
\label{fig:prof06}
\end{figure*}

\subsection{A Case Study}
\label{sec:case}

Before addressing how the evolution of the bound fraction,
$\fbound(t)$, depends on the orbital radius and on various numerical
parameters, we first examine one simulation in detail.
Fig.~\ref{fig:example} shows nine snapshots of one of our fiducial 
simulations (see Table~1), which
follows the evolution of a subhalo on a circular orbit with $\rorbNorm
= 0.5$. Each snapshot, the bound and unbound particles are indicated
by red and black dots, respectively.  The blue, dashed circle
indicates the initial virial radius of the subhalo, which coincides
with the initial truncation radius. The green, dashed circle marks the
initial orbit on which the subhalo is placed. The upper left-hand
corner of each panel indicates the time (in Gyr) and the bound
fraction, $\fbound$, while the inset shows a zoom-in on the central
region of the subhalo. The orbital time at $\rorbNorm = 0.5$ is
$T_{\rm orb} = 5.9 \Gyr$, so that at the end of the simulation
($t=63.5\Gyr$) the subhalo has almost completed $11$ orbits around the
centre of its host.

During the first orbital period, the subhalo looses about 68 percent
of its mass, which increases to 75 percent after 10 Gyr.  During the
subsequent 53 Gyr, the subhalo only looses an additional $\sim 13$
percent, while the tidally stripped material from the leading and
trailing tidal arms is phase-mixed into two ring-like features. The
surviving remnant, which after 60 Gyr still contains over $11,000$
particles continues to orbit in between these two `rings'. Due to the
perpetual sequence of stripping and re-virialization the subhalo
continues to loose mass, even after 60 Gyr, albeit at a very slow
rate.

Fig.~\ref{fig:prof06} plots the density profiles (left-hand panel),
circular velocity profiles (middle panel) and the radial velocity
dispersion profiles (right-hand panels) of the bound subhalo in
Fig.~\ref{fig:example}. Thin (green) and thick (black) lines show the
profiles every 2.5 and 25Gyr, respectively. Note how the density
profile in the central region remains fairly constant over time, while
the density profile in the outer regions becomes steeper and steeper
as more and more matter is being stripped. After 50 Gyr, when the
bound fraction has dropped to $\fbound = 0.122$, the mass enclosed
within ten percent of the original scale radius has only been reduced
by $\sim 40$ percent, while the maximum (radial) velocity dispersion
has dropped by $\sim 35$ percent. Overall these results are in good
agreement with previous studies \citep[e.g.,][]{Hayashi.etal.03,
  Penarrubia.etal.10}.

\begin{figure}
\includegraphics[width=\hssize]{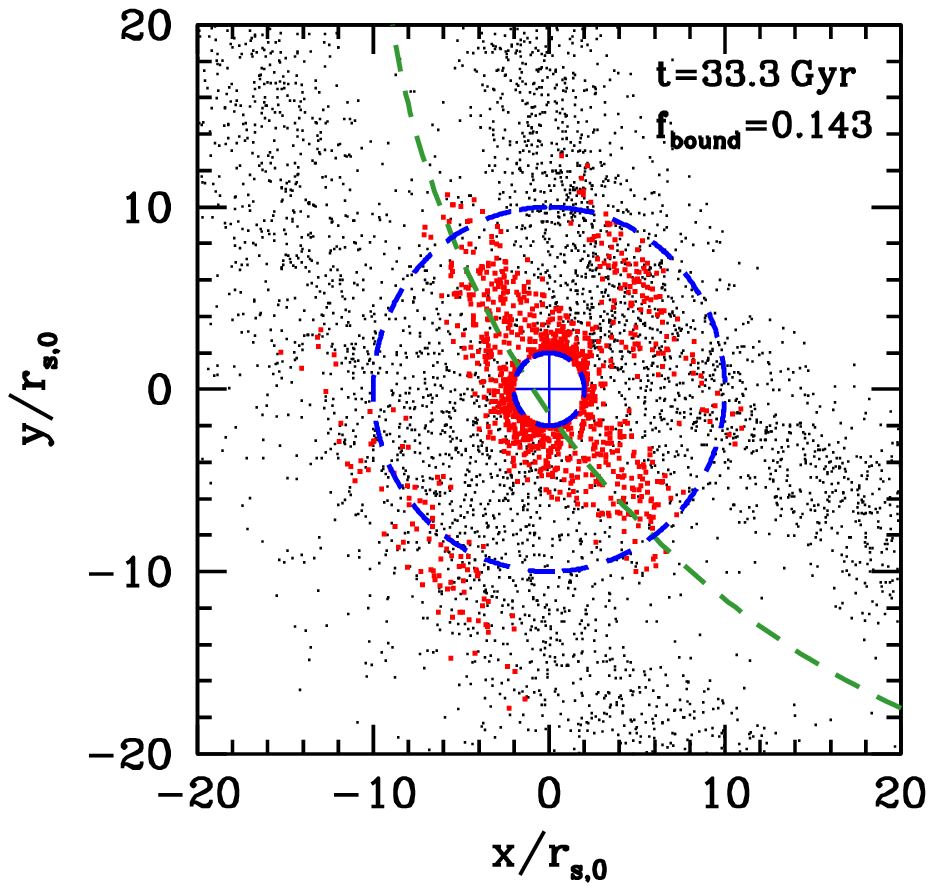}
\caption{A snapshot of the subhalo in our fiducial simulation of
  Fig.~\ref{fig:example} at $t=33.3 \Gyr$. Bound and unbound
  particles are indicated by red and black dots, respectively. For
  enhanced clarity, we only show the particles in the subhalo
  outskirts ($r > 2 \rs$; the region inside of this is marked with a
  crosshair). As in Fig.~\ref{fig:example} the blue and green dashed
  circles indicate the subhalo's original virial radius and orbit,
  respectively. Note the strongly anisotropic distribution of bound
  particles in the subhalo's outskirts (see text for discussion).}
\label{fig:zoom}
\end{figure}

The cautious reader may have noticed a pronounced, secondary `bump' in
the velocity dispersion profiles (right-hand panel of
Fig.~\ref{fig:prof06}) at late times ($t \gta 10 \Gyr$). This is due
to material that is in the act of being stripped, with a small
contribution due to transient phenomena. This is evident from
Fig.~\ref{fig:zoom}, which shows a zoom-in on the subhalo at $t=33.3
\Gyr$, where, for clarity, the central region with $r < 2 \rs$ is
cut-out. As in Fig.~\ref{fig:example}, red and black dots show the
bound and unbound particles, respectively, while the blue-dashed
circle marks the original virial radius of the subhalo.  Note that the
bound particles in the subhalo's outskirts are distributed very
anisotropically. There is a band of bound particles that fall roughly
along the subhalo's orbit (indicated by the green-dashed
circle-segment), and two `shells' of particles that are roughly
parallel to this band. The former are particles that have recently
been `stripped' from the subhalo in that they are clearly tracing out
a tidal stream. Formally, however, they are still bound to the
subhalo, in the sense that they pass the boundness criterion adopted
here (Eq.~[\ref{Ebound}]). It are these particles, which typically
become unbound within a small fraction of the subhalo's orbital
period, that are responsible for the secondary bump in the velocity
dispersion profile. The two shells of `bound' particles are made up of
subhalo particles that were stripped at much earlier times, and that
temporarily happen to pass the boundness criterion
(Eq.~[\ref{Ebound}]). We have verified that this transient bound
population is small, and does not significantly impact our estimates
of $f_{\rm bound}(t)$.

\subsection{Dynamical Self-Friction}
\label{sec:dynfric}

Upon closer inspection, the insets in the lower panels of
Fig.~\ref{fig:example} show that, at late times, the bound remnant of
the subhalo is orbiting inside of its original orbit (indicated by the
green, dashed circle). This is a manifestation of `dynamical
self-friction' in which the stripped material exerts a force on the
bound remnant, causing it to loose specific orbital energy to the
tidally stripped material \citep[see also][]{Fujii.etal.06,
  Fellhauer.Lin.07}. Note that dynamical friction due to the host halo
is absent in our simulations, since the host halo is modeled using an
analytical, static potential.

A detailed study of this dynamical self-friction will be presented in
a forthcoming paper (van den Bosch \& Go, in preparation).  In the
case of the simulations presented in this paper, which all have
$\ms/M_\rmh = 1/1000$, the impact of dynamical self-friction is
relatively weak, only affecting the orbital radius by a few percent at
most. However, for larger mass ratios, $\ms/M_\rmh$, (not treated
in this paper) we find the effect to be significantly larger, to the
extent that it can cause the subhalo-remnant to loose most of its
orbital energy and angular momentum in much less than a Hubble time.
In general, dynamical self-friction transports the remnant towards the
centre of the host halo, where the tidal field is stronger. This
causes a decrease in the tidal radius and thus an enhancement in the
mass loss rate. Hence, in addition to re-virialization, self-friction
adds another positive feedback loop to the non-linear behavior of
$\fbound(t)$, further inhibiting an analytical treatment.

\subsection{Dependence on orbital radius}
\label{sec:rorb}

\begin{figure}
\includegraphics[width=\hssize]{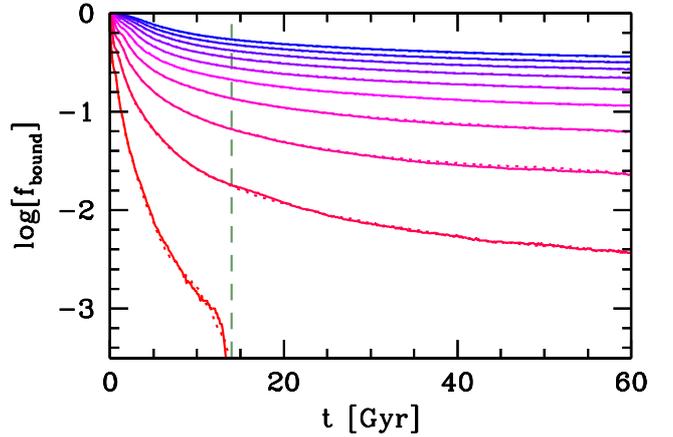}
\caption{The bound fraction of subhaloes as function of time (in
  $\Gyr$). Colors from blue to red correspond to $\rorbNorm = 1.0,
  0.9,...,0.1$, while solid and dotted lines (which almost lie on top
  of each other) correspond to simulations run with \treecode and
  \GPUtree, respectively. The dashed, vertical line marks the Hubble
  time, $t_\rmH = 13.97$ Gyr. Note how the results from the two
  different simulation codes are in excellent agreement with each
  other, and how all subhaloes survive for $> 60\Gyr$, except for the
  subhalo on a circular orbit with $r_{\rm orb} = 0.1 r_{\rm vir,h}$,
  which disrupts after $\sim 13\Gyr$.}
\label{fig:circular}
\end{figure}

The colored lines in the upper left-hand panel of
Fig.~\ref{fig:circular} show the bound fraction as function of time
for subhaloes on different circular orbits, with $\rorbNorm$ ranging
from $1.0$ (blue line) to $0.1$ (red line). All these simulations are
run using the fiducial parameters listed in Table~1.  Solid and dotted
lines correspond to simulations run with \treecode and \GPUtree,
respectively, and are in excellent agreement, indicating that our
results are independent of which of the two codes we use.  Note how
all subhaloes survive for the full 60 Gyr of the simulation, except
for the subhalo on the orbit with $\rorbNorm = 0.1$, which disrupts
after $\sim 13\Gyr$ (corresponding to $\sim 8.5$ orbits). Hence, based
on these simulations one might be tempted to conclude that for the
present set-up ($M_\rmh/\ms = 1000$, $c_\rms = 10$, $c_\rmh = 5$, and
circular orbits), subhaloes disrupt when $\rorbNorm \lta
0.15$. However, as we demonstrate below, there are a number of
numerical issues that have a strong impact on the results shown.

\section{Numerical Issues}
\label{sec:num}

One of the main goals of this paper is to assess to what extent
numerical $N$-body simulations are able to resolve the dynamics (i.e.,
accurately integrate the equations of motion) related to the tidal 
evolution of dark matter substructure. In this section we assess the 
impact of four numerical parameters that control the accuracy of such 
simulations: the time step, $\Delta t$, the tree opening angle, $\theta$, 
the softening length, $\varepsilon$, and the actual number of particles,
$\Np$, used to simulate the system in question.

\subsection{Timestepping}
\label{sec:timestep}

An important requirement for numerical $N$-body simulations is that the
equations of motion are integrated accurately. In general, accuracy
improves when using smaller time steps, but at the expense of increased
computational cost. Modern cosmological simulations typically use
adaptive time stepping, in which different criteria are used to
determine the time step for each individual particle.  However,
adaptive time stepping is non-trivial and faces numerous challenges
\citep[see][for an overview]{Dehnen.Read.11}.  In this paper, we are
conservative and only use simulations that employ a fixed time step,
$\Delta t$, for all particles.
\begin{figure*}
\includegraphics[width=0.9\hdsize]{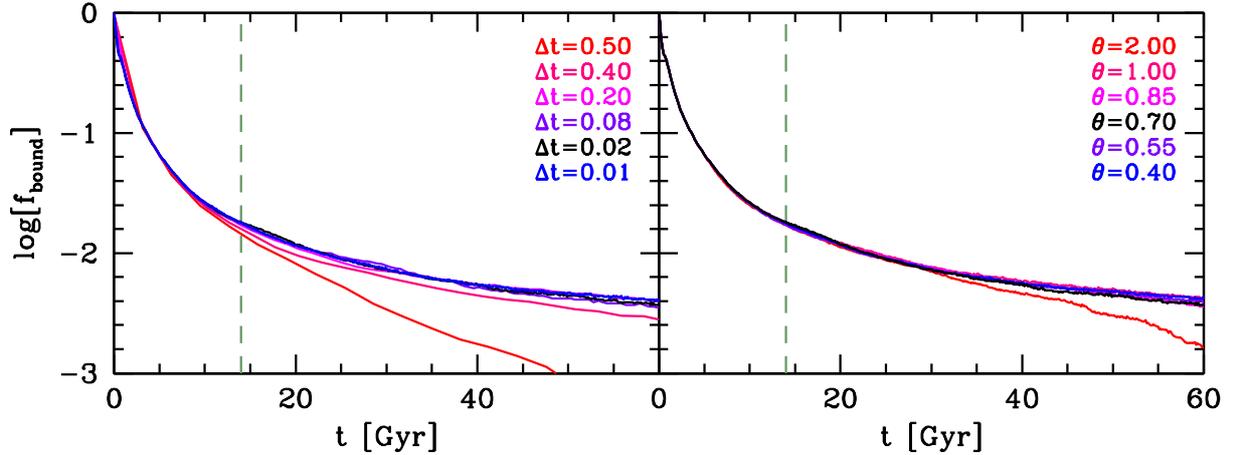}
\caption{Impact of changing the time-step $\Delta t$ (left-hand
  panel) and the opening angle $\theta$ (right-hand panel) on the
  evolution of $\fbound(t)$ for simulations with $\rorbNorm=0.2$. All
  simulations use $\Np=10^5$ and a softening length
  $\varepsilon=0.05$. The black line corresponds to our fiducial
  simulation, which has $\theta=0.7$ and $\Delta t = 0.02$, and the
  vertical dashed line indicates the Hubble time.}
\label{fig:NumPar}
\end{figure*}

It is important to choose $\Delta t$ sufficiently small, such that one
accurately resolves the dynamics of the system. We require that
$\Delta t$ be significantly smaller than the {\it minimal} orbital time
of the system, defined as the orbital time at the radius equal to the
softening length, $\tau_{\rm min} = [3 \pi / G
  \bar{\rho}(<\varepsilon)]^{1/2}$, with $\bar{\rho}(<\varepsilon)$
the average density enclosed by the softening length. In particular,
we adopt $\Delta t \leq \tau_{\rm min}/30$. The subhaloes simulated 
here have initial NFW density profiles with $c=10$, for which
\begin{equation}\label{dtcrit}
\Delta t < 0.256 \sqrt{ \frac{\varepsilon^3}{\ln(1+\varepsilon) - 
  \frac{\varepsilon}{1+\varepsilon}}}\,.
\end{equation}
For the range of softening parameters considered in this paper ($0.003
< \varepsilon < 0.25$), this implies $0.02 < \Delta t < 0.21$, with
$\Delta t = 0.08$ for our fiducial set-up ($\varepsilon=0.05$, see
\S\ref{sec:softening}). 

In addition to properly resolving the relevant orbital times, 
the time step also impacts the accuracy of close encounters among particles.
If the time step is too large, the resulting errors in the integration of
such encounters causes an exacerbation of discreteness effects. As discussed 
in \cite{Power.etal.03}, achieving convergence down to a particular radius 
of a dark matter halo, for a particular duration of integration, puts a 
constraint on the ratio $\Delta t / (\Np \varepsilon)$; for smaller ratios, 
the radius out to which the halo is properly resolved is smaller. 

To test how our simulation results depend on the choice of time step,
we have performed a series of simulations with different $\Delta t$.
The left-hand panel of Fig.~\ref{fig:NumPar} shows the evolution of
the bound mass fraction of our fiducial subhalo on a circular orbit
with $r_{\rm orb} = 0.2 r_{\rm vir,h}$. Different colors correspond to
different time steps, as indicated, while all other parameters are
kept fixed at their fiducial values ($\Np=10^5$, $\varepsilon=0.05$,
and $\theta=0.7$). The results are nicely converged as long as
$\Delta t \lta 0.4$. Hence, based on the study by \cite{Power.etal.03} 
we conclude that properly resolving the tidal evolution of dark matter 
substructure requires a time step
\begin{equation}
\Delta t < (\Delta t)_{\rm max} = 0.4 \left(\frac{\Np}{10^5}\right) \,
\left(\frac{\varepsilon}{0.05}\right)
\end{equation}
We have repeated this test for other settings (other $\Np$, $\varepsilon$ and
$r_{\rm orb}/r_{\rm vir,h}$), and find that the above criterion accurately
delineates the boundary between converged simulations and simulations in which 
the time step is too small. The latter manifests in excessive mass loss and 
premature disruption, compared to simulations with $\Delta t < (\Delta t)_{\rm max}$.
In what follows we are conservative and always adopt a time step
$0.02 \leq \Delta t \leq (\Delta t)_{\rm max}$, which assures that 
accurate time integration of the equations of motion is not a limiting 
factor in any of the simulations presented below.

\subsection{Force Accuracy}
\label{sec:theta}

Both simulation codes used here are tree codes, which make use of the
fact that the gravitational potential of a distant group of particles
can be well-approximated by a low-order multipole expansion. In a tree
code, the particles are therefore arranged in a hierarchical system of
groups (cells) that form a tree structure.  Forces are evaluated by
`walking' down the tree level by level, beginning with the top
cell. At each level, a cell is added to an interaction list if it is
distant enough for a force evaluation; if the cell is too close, it is
`opened' and the subcells (of which there are 8 in the case of an
octree) are either used for force evaluation or opened further. A cell
is opened, whenever $d < l/\theta + \delta r$.  Here $d$ is the
distance between the particle in question and the centre-of-mass of
the cell, $\delta r$ is the distance between the centre of the cell
and its centre-of-mass, and $\theta$ is the opening angle, which
controls the accuracy of the force calculation. Smaller $\theta$
results in more accurate forces, but also increases computational
cost.
\begin{figure*}
\includegraphics[width=0.98\hdsize]{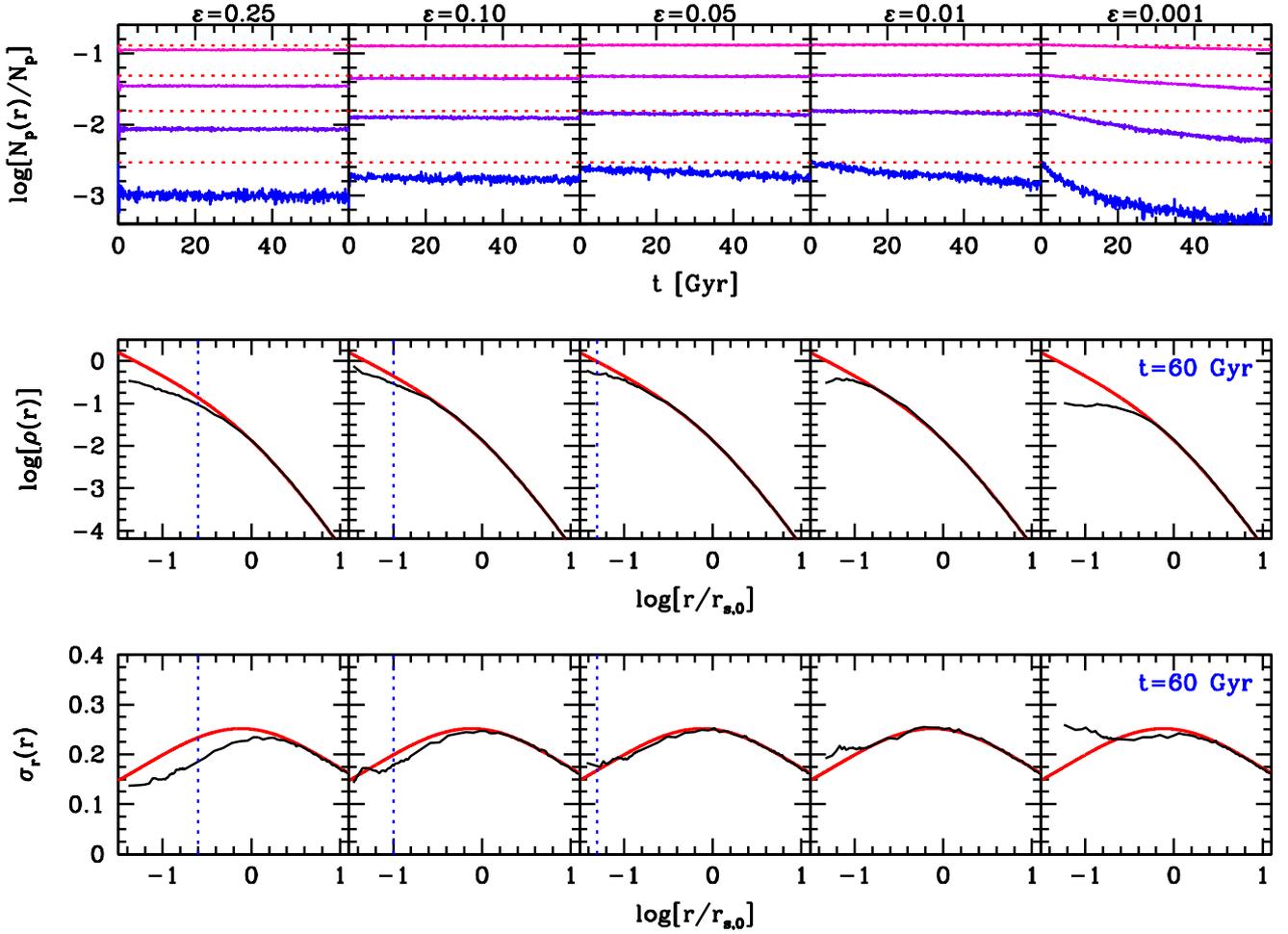}
\caption{Impact of force softening on the evolution of an isolated NFW
  halo with $N=10^5$ particles inside its virial radius. Different
  columns correspond to different values of the softening length,
  $\varepsilon$, as indicated (in model units) at the top of each
  column. Upper panels show the normalized number of particles,
  $\Np(r)/\Np$, enclosed within fixed radii of $r/r_{\rm vir} = 0.01$,
  $0.025$, $0.05$ and $1.0$ (different colors).  The red, dotted,
  horizontal lines indicate the corresponding analytical predictions
  (i.e., in the absence of softening and in the limit of $\Np
  \rightarrow \infty$).  Panels in the middle and lower rows show the
  density profile, $\rho(r)$, and radial velocity dispersion profile,
  $\sigma_r(r)$, after $t=60\,$Gyr, compared to their respective
  analytical profiles at $t=0$ (thick red lines).  The vertical dashed
  lines indicate the softening length. See text for a detailed
  discussion.}
\label{fig:softening}
\end{figure*}

Throughout we set $\theta = 0.7$, and we have verified through a
number of tests that our results are extremely robust to changes in
$\theta$.  One such test is shown in the right-hand panel of
Fig.~\ref{fig:NumPar}, which plots the evolution of the bound mass
fraction of our fiducial subhalo on a circular orbit with $r_{\rm orb}
= 0.2 r_{\rm vir,h}$. Different colors correspond to different values
of $\theta$, as indicated, while all other parameters are kept fixed
to their fiducial values (i.e., $\Np=10^5$, $\varepsilon=0.05$, and
$\Delta t = 0.02$). Note that decreasing $\theta$ to $0.4$ does not
significantly change the results, indicating that our simulations are
not limited by force accuracy. Only when we adopt extremely large
values for the opening angle (i.e., $\theta \gta 2$) do we notice
significant departures from the converged (i.e., for much smaller
$\theta$) results. Throughout we are conservative and always adopt
$\theta=0.7$.
\begin{figure*}
\includegraphics[width=0.9\hdsize]{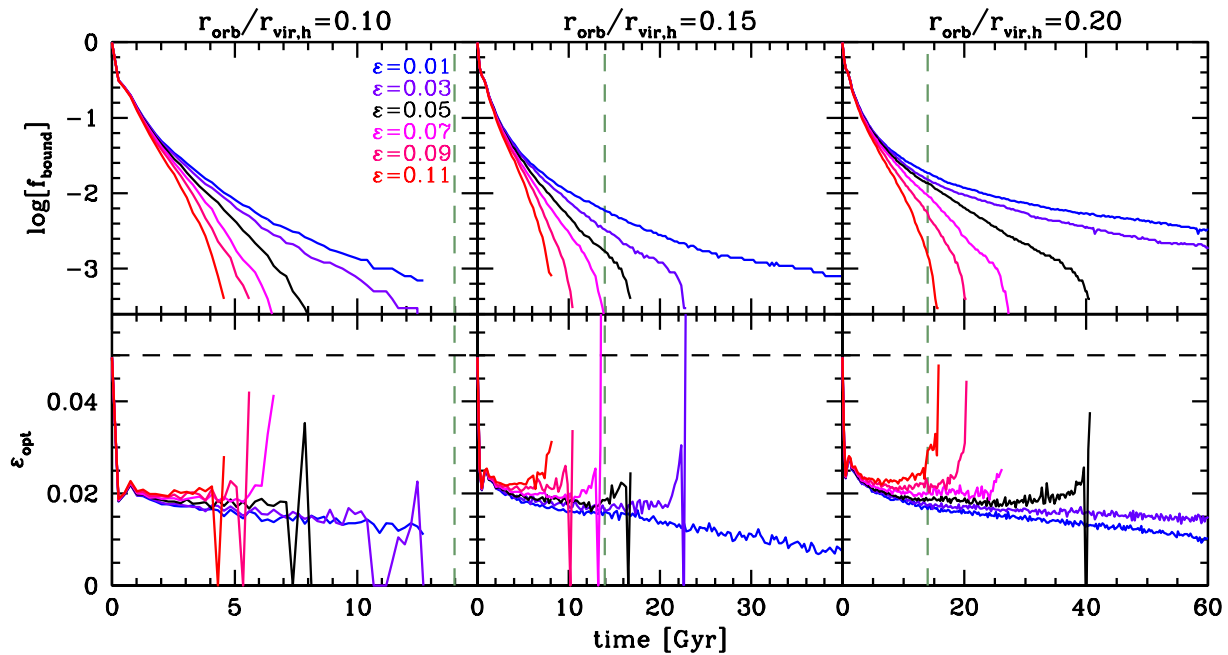}
\caption{Impact of force softening on the evolution of a subhalo on a
    circular orbit in the tidal field of a fixed, analytical host
    halo. Top and bottom panels show the evolution of the bound mass
    fraction, and the value for the optimal softening length,
    respectively.  The latter is defined as $\varepsilon_{\rm opt} =
    0.05 (r_\rmh/\rh) \fbound^{-1/3}$. Lines of different color
    correspond to different values for the softening length, while
    different columns correspond to different orbital radii, as
    indicated.  As in previous figures, the dashed, vertical lines
    indicate the Hubble time. Note how decreasing the softening length
    increases the bound mass fraction, causing the subhalo to survive
    longer.}
  \label{fig:soft}
\end{figure*}

\subsection{Force Softening}
\label{sec:softening}

Typically, the choice for the softening length, $\varepsilon$, is a
compromise between too much force {\it bias} (if $\varepsilon$ is too
large) and too much force {\it noise} (if $\varepsilon$ is too
small). Whereas the former results in a central potential well that is
too shallow, the latter results in too many large-angle scattering
events that create an artificial, isothermal core
\citep[e.g.,][]{Athanassoula.etal.00, Dehnen.01}.  Based on these
considerations, \cite{Dehnen.01} advocates that a Hernquist sphere
(which has the same $r^{-1}$ cusp as an NFW sphere), has an optimal
softening length equal to $\varepsilon_{\rm opt}/\rs = 0.017 \,
N_5^{-0.23}$, where $N_5 = \Np/10^5$.  \cite{Power.etal.03} suggests
picking a softening length for which the maximum stochastic
acceleration caused by close encounters with individual particles,
$a_{\rm stoch} \simeq Gm_\rmp/\varepsilon^2$, be smaller than the
minimum mean-field acceleration in the halo, $a_{\rm min} \simeq G
M_{\rm vir}/r^2_{\rm vir}$. This implies $\varepsilon_{\rm opt} \simeq
r_{\rm vir}/\sqrt{N_{\rm vir}}$, which translates to $\varepsilon_{\rm
  opt}/\rs \simeq 0.032 \, (c_\rms/10) \, N_5^{-1/2}$. Note that
\cite{Power.etal.03} actually advocate using a softening that is a
factor four times larger, but this is based on a trade-off between
accuracy and minimizing the number of time steps required to reach
convergence down to some radius. Since we are using relatively small
simulations, minimizing the number of time steps is not a major
concern to us, and we therefore opt for the smaller, more conservative
optimal softening length. Finally, \cite{vKampen.00a} advocates yet
another criterion for the softening length, based on demanding
(roughly) that the mean particle separation within the half-mass
radius, $r_\rmh$, be larger than the softening length. This implies
$\varepsilon_{\rm opt} \propto 0.77 \, r_\rmh \, \Np^{-1/3}$. For an
NFW halo, we find that to good approximation
\begin{equation}\label{rhrs}
\frac{r_\rmh}{r_\rms} = 3.6 \times \left(\frac{c}{10}\right)^{0.63}
\end{equation}
which implies $\varepsilon_{\rm opt}/\rs \simeq 0.060 \,
(c_\rms/10)^{0.63} \, N_5^{-1/3}$.

Based on the three criteria discussed above, the optimal softening
length (in model units) for our fiducial simulations, which have $N_5
= 1$ and $c_\rms=10$, ranges from 0.02 to 0.06. To test this, we have
performed a set of simulations of a NFW halo in isolation (i.e., no
external tidal field), in which we only vary the softening parameter.
Each simulation models the evolution of an NFW halo with concentration
$c=10$ out to $\rmax = 10\,\rvir$. We simulate the system with $\Np =
2.43 \times 10^5$ particles in total, such that the number of
particles inside the virial radius equals $10^5$. We adopt our
fiducial time step ($\Delta t = 0.02$) and opening angle ($\theta =
0.7$), and vary $\varepsilon$ from 0.001 to 0.25. Each simulation is
run for 50.000 time steps (corresponding to $\sim 63\,$Gyr). The
results are shown in Fig.~\ref{fig:softening}, where the upper panels
show the normalized number of particles enclosed within fixed radii of
$r/r_{\rm vir} = 0.01$, $0.025$, $0.05$ and $1.0$ (different colors).
The red, dotted, horizontal lines indicate the analytical predictions
(i.e., in the absence of softening and in the limit of $\Np
\rightarrow \infty$).  Panels in the middle and lower rows show the
halo density profile and halo velocity dispersion profile,
$\sigma(r)$, after $t=60\,$Gyr, compared to their respective
analytical profiles at $t=0$ (thick red lines).

As is evident, the results are `optimal' when $\varepsilon \simeq
0.05$, in good agreement with the various $\varepsilon_{\rm
  opt}$-criteria above.  When $\varepsilon$ is too large, the halo
potential is overly softened, resulting in the density and velocity
dispersion being too small at small radii. Since less dense systems
are more easily stripped, we expect that in simulations with
$\varepsilon > \varepsilon_{\rm opt}$ subhaloes experience enhanced
stripping, and possibly artificial disruption.  When $\varepsilon$ is
too small, the enclosed mass evolves with time due to two-body
relaxation effects; energy is transported inwards, causing $\sigma(r)$
to become isothermal in the centre, and the central density profile to
evolve from a cusp into a core. Hence, the naive expectation is that
also for $\varepsilon < \varepsilon_{\rm opt}$ subhaloes experience
enhanced tidal stripping and/or disruption.

To test how softening impacts the tidal evolution of subhaloes, we now
put our fiducial NFW halo on a circular orbit within the (static)
tidal field of our fiducial host halo ($M_\rmh=1000,
c_\rmh=5$). Fig.~\ref{fig:soft} show the results for three such orbits
($r_{\rm orb}/r_{\rm vir,h} = 0.1$, $0.15$ and $0.2$, different
columns), and for 6 different softening parameters each ($\varepsilon
= 0.01, 0.03,...,0.11$; different colors). The upper panels show the
evolution of the bound mass fraction with time.  As expected, when
$\varepsilon > \varepsilon_{\rm opt} \simeq 0.05$, subhaloes
experience enhanced stripping and disruption. However, contrary to
expectations, when $\varepsilon < \varepsilon_{\rm opt}$ the simulated
subhaloes have {\it larger} bound fractions and survive {\it longer}.
The reason for this counter-intuitive behavior is that the optimal
softening length scales with both the size of the system and the
number of (bound) particles, both of which are evolving rapidly. 
The lower panels of Fig.~\ref{fig:soft} plot the evolution of the 
optimal softening length, based on the scaling relation advocated by
\cite{vKampen.00a}; $\varepsilon_{\rm opt} = 0.05 (r_\rmh/r_{\rmh,0}) 
f^{-1/3}_{\rm bound}$, where $r_{\rmh,0}$ is the initial half-mass 
radius (prior to stripping). Note how $\varepsilon_{\rm opt}$ rapidly 
drops to $\sim 0.02$, after which it slowly decreases to $0.01$. Most 
interestingly, all simulations show roughly the same behavior, independent 
of the actual softening length used. This suggests that the optimal softening
length for subhaloes is smaller than that for isolated host haloes and
that $\varepsilon_{\rm opt}$ may actually depend on the strength of
the tidal field. We will come back to this issue in
\S\ref{sec:convergence}, but for now we conclude that changes in the
softening parameter has a drastic impact on the tidal evolution of
substructure in numerical simulations.
\begin{figure*}
\includegraphics[width=\hdsize]{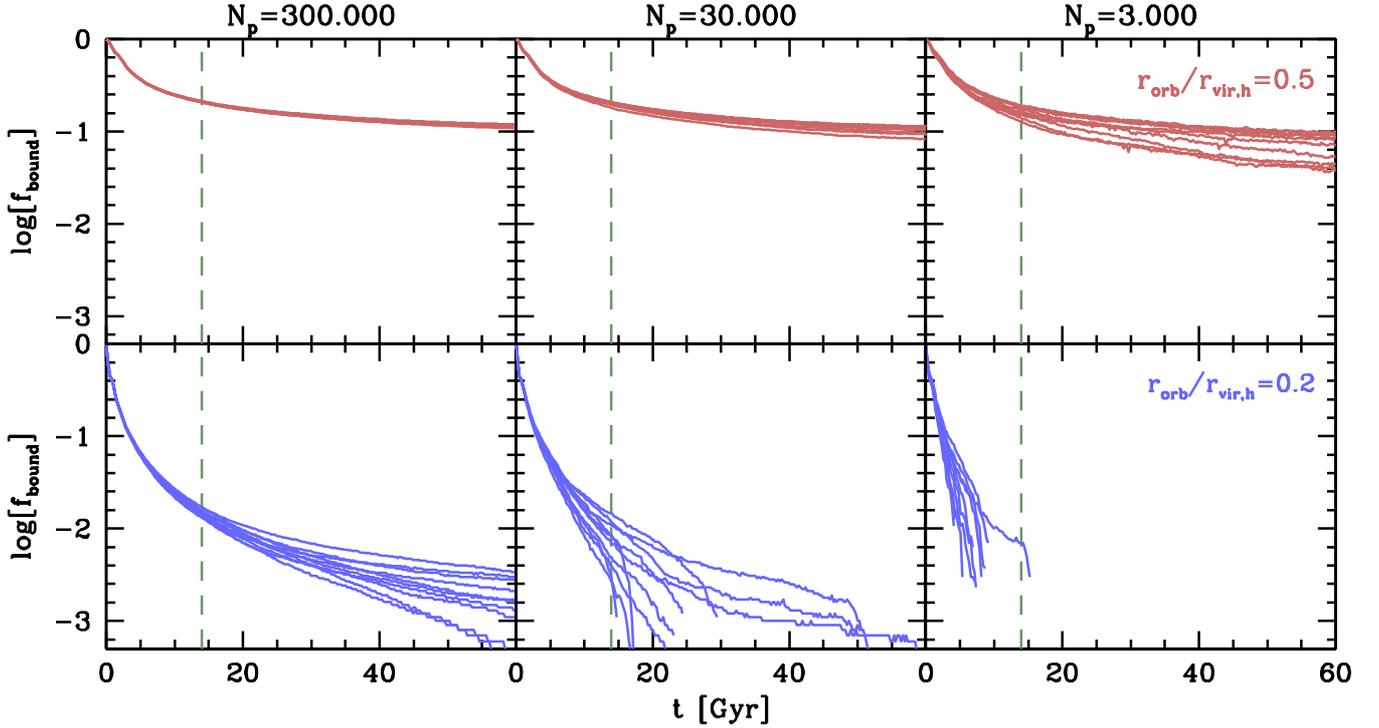}
\caption{The bound mass fraction as function of time. Each panels
  shows 10 independent random realizations for a subhalo on a circular
  orbit with $\rorbNorm=0.5$ (upper panels) or $\rorbNorm=0.2$ (lower
  panels). Results are shown for simulations with $\Np=300,000$
  (left-hand column), $\Np=30,000$ (middle column) and $\Np=3,000$
  (right-hand column) particles, and a softening length of
  $\varepsilon = 0.05 \,(\Np/10^5)^{-1/3}$. The vertical dashed line
  indicates the Hubble time. Note how the variance among the different
  random realizations increases with time due to a discreteness-driven
  runaway instability.}
\label{fig:Ndep}
\end{figure*}

\subsection{Discreteness Noise}
\label{sec:discreteness}

When using a fairly limited number of particles to sample the DF, one
is subject to discreteness noise. The corresponding Poisson
fluctuations in the density and potential of the subhalo impact the
dynamics. For instance, they cause collective relaxation (a numerical
artefact), which is reminiscent of violent relaxation (a physical
process), and which, under certain conditions, can dominate over
two-body relaxation \citep{Weinberg.93}.  To gauge the impact of
discreteness noise on the evolution of $\fbound(t)$, we have run a
large number of simulations using different random realizations when
setting up the ICs. The findings are summarized in Fig.~\ref{fig:Ndep}
which plots the bound mass fraction as function of time. Results are
shown for two different orbits, $\rorbNorm = 0.5$ (upper panels) and
$\rorbNorm = 0.2$ (lower panels), and for three different mass resolutions,
$\Np=300,000$ (left-hand panels), $\Np=30,000$ (middle panels) and
$\Np=3,000$ (right-hand panels). When changing the number of
particles, we also change the softening length according to
$\varepsilon = 0.05 (\Np/10^5)^{-1/3}$ (see \S\ref{sec:numsim}).

Each panel shows the results for ten different simulations that only
differ in the random realization of the ICs. In the absence of
discreteness noise these ten realizations should yield exactly
identical results. In practice, however, the simulations display
diverging behavior in $\fbound(t)$, which becomes more pronounced for
smaller $\Np$ and/or smaller $\rorbNorm$. In some cases, the
divergence is dramatic.  For example, in the case with $\rorbNorm=0.2$
and $\Np=30,000$ some subhaloes survive for $> 60 \Gyr$, while others
disrupt in just under a Hubble time. When $\Np=3,000$ the subhalo
always disrupts, but with a disruption time that varies from $\sim
5\Gyr$ to $\sim 15 \Gyr$.  Even with $\Np=300,000$, the discreteness
noise results in a simulation-to-simulation variance in $\fbound$
after $60 \Gyr$ with a standard deviation of $\sim 0.4$ dex.

\begin{figure}
\includegraphics[width=\hssize]{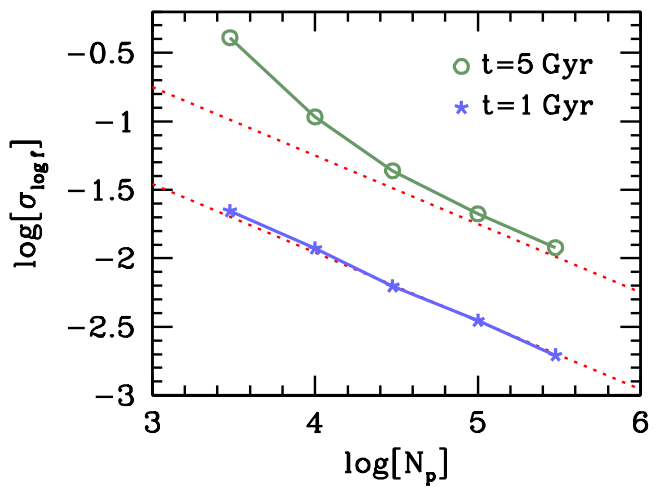}
\caption{The standard deviation in $\log \fbound$ among 10 different
  realizations as a function of the logarithm of the number of
  particles used in the simulations.  Results are shown at two epochs,
  1 and 5 Gyrs, after the start of the simulations, as indicated. The
  red, dotted lines correspond to $\sigma_{\log f} \propto \Np^{-1/2}
  \, t$, and are shown for comparison.  Note that increasing $\Np$
  results in a clear reduction of the standard deviation among
  different realizations, indicating that the variance is mainly due
  to discreteness noise. All these results are for simulations with
  $\rorbNorm=0.2$ and adopt a softening length of $\epsilon = 0.05 \, 
  (N_{\rm p}/10^5)^{-1/3}$.}
\label{fig:siglogf}
\end{figure}

We can characterize the sensitivity to this discreteness noise using
the variance, $\sigma_{\log f}$, in $\log(\fbound)$ among these sets
of simulations. Fig.~\ref{fig:siglogf} plots the logarithm of
$\sigma_{\log f}$ in the $\rorbNorm=0.2$ simulations as a function of
$\log(\Np)$. Results are shown for two epochs, 1 and 5 Gyrs after the
start of the simulations. As expected, the standard deviation
increases with time and with decreasing number of particles.  The
red-dotted lines in Fig.~\ref{fig:siglogf} correspond to
\begin{equation}\label{NoiseScaling}
\sigma_{\log f} \propto \Np^{-1/2} \, t\,,
\end{equation}
anchored to the standard deviation at $\log \Np = 5$ and $t =
1\Gyr$. Clearly, this scaling, which is in line with simple
predictions for discreteness noise, provides a fairly good description
of the simulation results. However, at later times, and small $\Np$,
the standard deviation is larger than predicted by
Eq.~(\ref{NoiseScaling}). We have performed a number of tests, mainly
varying softening lengths, which suggest that this deviation arises
because for small $\Np$ the dynamics at later times is strongly
influenced by force bias arising from poor softening: when using
smaller softening lengths the simulation results more closely follow
the scaling relation of Eq.~(\ref{NoiseScaling}). We have also
repeated the above experiments in which we first run the haloes in
isolation for $10 \Gyr$, allowing them to virialize before we
introduce them to the tidal field of the host halo. The resulting
divergence is indistinguishable from that in the runs shown above.
This indicates that the diverging behavior among the various 
realizations is due to discreteness rather than due to the initial 
conditions not being in perfect equilibrium (see \S\ref{sec:ICsens} 
for more details).

Although the simulations roughly obey the scaling relation of
Eq.~(\ref{NoiseScaling}), we emphasize that its {\it normalization}
depends on a number of physical parameters such as the tidal field
strength (e.g., $r_{\rm orb}$, $M_\rmh$, $c_\rmh$) and the structural 
parameters of the subhalo (i.e., $c_\rms$).  Typically, the impact of
discreteness noise on the evolution of the subhalo increases with the 
strength of the tidal field. This is evident from comparing the upper 
and lower panels of Fig.~\ref{fig:Ndep}, which show a much stronger 
divergence in $\fbound$ for smaller $\rorb$. In addition, we have 
run the same ten realizations in isolation (i.e., in the absence of 
an external tidal field), in which case we find no significant 
divergence in their evolution.

\subsection{A Discreteness-Driven Run-Away Instability}
\label{sec:runaway}

In order to understand the origin of the divergence in the bound mass 
fractions, it is useful to envision the tidal radius as characterizing
a semi-permeable surface; once a particle crosses
this surface it will be stripped off. Due to discreteness noise, the
number of particles that cross this tidal surface during any time
interval $\Delta t$ is subject to Poisson fluctuations.  Now consider
a high fluctuation, i.e., a time interval during which the mass loss
rate is higher than average. As a consequence, the remaining remnant
experiences more re-virialization than average, which results in its
size expanding more than average. As a result, the tidal radius
shrinks more than average, and the next time step $\Delta t$ the
subhalo is therefore more likely to once again loose more mass than on
average. Hence, discreteness noise gives rise to a run-away
instability that is triggered by the presence of a tidal field. This
is an important, and potentially worrying result, as it implies that
the evolution of subhaloes in cosmological simulations is likely to be
severely impacted by discreteness noise. As the results shown here
suggest, this discreteness noise can trigger artificial disruption of
subhaloes even when they are still resolved with thousands of
particles.

\section{Towards Convergence}
\label{sec:convergence}

In the previous section we have demonstrated that our simulation
results are subject to a run-away instability due to discreteness
noise and extremely sensitive to the softening length used, whose
optimal value appears to depend on the strength of the tidal field. 
This begs the question; what is the correct tidal evolution of
substructure and what are the numerical requirements to properly track
such evolution in $N$-body simulations?  Unfortunately, as discussed
in \S\ref{sec:stripping} and Paper~I, there is no analytical answer,
and we therefore have to rely on numerical simulations. We now
undertake a detailed resolution study, in which we increase $\Np$ and
vary $\varepsilon$ until we reach a `converged' $f_{\rm
  bound}(t)$. Here converged means that (i) no significant changes
occur when $\Np$ is increased further\footnote{while 
decreasing the softening accordingly, see \S\ref{sec:softening}}, and 
(ii) the standard deviation in $f_{\rm bound}$ after one Hubble time 
is sufficiently small (i.e., $\sigma_{\log f} \lta 0.05$).
\begin{figure*}
\includegraphics[width=0.95\hdsize]{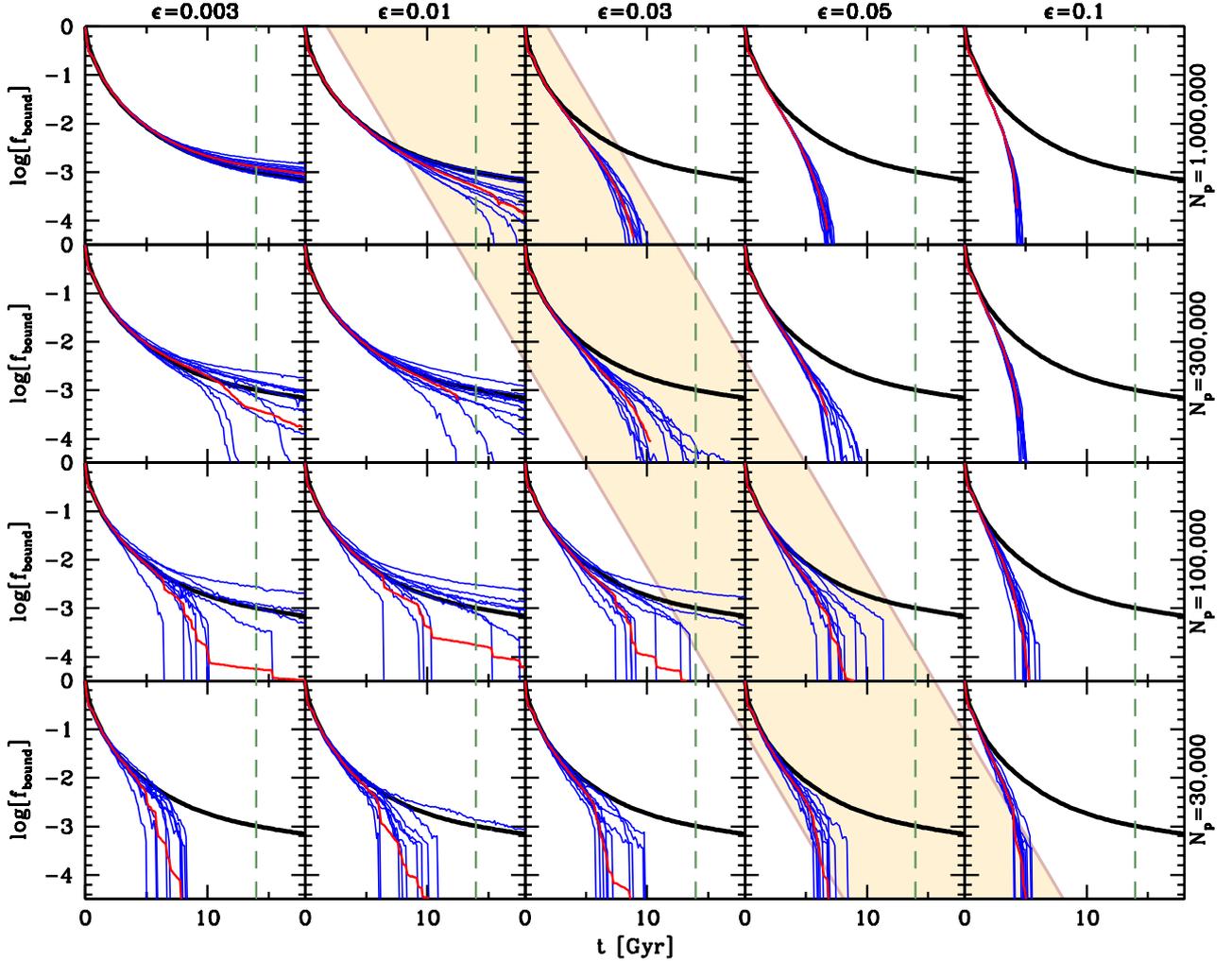}
\caption{Bound fraction as function of time for simulations with
  different $\Np$ (different rows) and different softening length,
  $\varepsilon$ (different columns). All simulations follow the
  evolution of a subhalo with $c_\rms=10$ on a circular orbit with
  $\rorbNorm=0.1$ in a host halo with $M_\rmh = 1000 \ms$ and $c_\rmh
  = 5$. The solid, black line indicates the converged results, based
  on a simulation with $\Np=10^7$ and $\varepsilon = 0.003$.  Blue
  lines indicate the results from 10 simulations, that only differ in
  the random realization of the initial conditions, while the red line
  indicates their average. Vertical, dashed lines marks the Hubble
  time, while the yellow shaded band indicates the typical
  $(\Np,\varepsilon)$ with which subhaloes are resolved in
  state-of-the-art cosmological simulations (the width of the band
  matches the width of the panels, and does not have a physical
  meaning). See text for a detailed discussion.}
\label{fig:r01}
\end{figure*}

We start by considering a circular orbit with $\rorbNorm = 0.1$.  As
shown in \cite{vdBosch.17}, this is the typical radius at which dark
matter subhaloes in the Bolshoi simulation \citep[][]{Klypin.etal.11}
undergo disruption, and by focusing on this orbit we therefore hope
to gain some understanding as to whether such disruption is real
(physical) or artificial (numerical). The results of our convergence
study for this orbit are shown in Fig.~\ref{fig:r01}, which summarizes
the results of over 200 simulations with different $\Np$ and
$\varepsilon$; all other parameters are kept fixed at their fiducial
values. Each panel shows the evolution of the bound fraction as
inferred from 10 random realizations (blue curves) with the same $\Np$
and $\varepsilon$. The red curve indicates the corresponding average
of $\log[f_{\rm bound}]$, while the thick solid curve indicates the
evolution of the bound fraction for the `converged' result, which in
this case has $\Np=10^7$ and $\varepsilon = 0.003$. From top to bottom
$\Np$ decreases from $10^6$ to $3\times 10^4$, while from left to
right $\varepsilon$ increases from $0.003$ to $0.1$, as indicated. The
vertical, dashed line marks the Hubble time.

The results in the upper-left corner ($\Np=10^6$ and
$\varepsilon=0.003$) are in good agreement with our converged results,
but has $\sigma_{\log f} \simeq 0.1$ (i.e., there is a $0.1$dex
uncertainty on the surviving halo mass) after one Hubble time.
Increasing the softening length while keeping the number of particles
fixed at $\Np=10^6$ causes a systematic deviation from the converged
(`true') results, the amplitude of which clearly increases with
$\varepsilon$; For $\varepsilon=0.03$, the subhaloes disrupt after $(9
\pm 1)$Gyr, while with $\varepsilon = 0.1$ the subhaloes disrupt after
only $\sim 5$Gyr.  If instead we keep $\varepsilon=0.003$ and decrease
$\Np$, the impact of discreteness noise becomes more and more
apparent. With $\Np=10^5$ the standard deviation $\sigma_{\log f}$
already reaches $0.1$dex after only $\sim 4$Gyr.

Clearly, properly resolving the dynamics associated with the tidal
evolution of substructure at 10 percent of the virial radius of the
host halo requires extremely high mass resolution (i.e., large $\Np$) as
well as superb force resolution (i.e., small $\varepsilon$).  To
compare this with the typical mass and force resolutions used in
state-of-the-art cosmological simulations, consider the Millennium
simulation \citep{Springel.etal.05}, which has a particle mass $m_\rmp
= 8.6 \times 10^8 \Msunh$ and a softening length of $5 h^{-1} \kpc$
comoving (Plummer equivalent). Using the concentration-mass relation
of \cite{Neto.etal.07}, according to which $c_{200} = 5.26
(M_{200}/10^{14}\Msunh)^{-0.1}$, this implies that haloes with
$N_{200}$ particles within the radius that encloses a density equal
to 200 times the critical density are resolved with a softening
length, in units of the NFW scale radius, given by
\begin{equation}\label{softscale}
\frac{\varepsilon}{r_\rms} = 0.037 \, \left(\frac{N_{200}}{10^5}\right)^{-0.43}\,,
\end{equation}
where we have used that $r_\rms = r_{200}/c_{200}$ and $r_{200}
\propto M_{200}^{1/3} \propto N_{200}^{1/3}$.  The yellow-shaded band
in Fig.~\ref{fig:r01} roughly reflects this scaling. It shows that the
Millennium simulation is unable to properly resolve the dynamical
evolution of subhaloes on circular orbits with $\rorbNorm \simeq 0.1$;
they are subject to the discreteness driven run-away instability
identified in \S\ref{sec:discreteness}, and they experience artificial
disruption due to a softening length that is too large.

It is important to emphasize that this is not specific to the
Millennium simulation. In fact, most state-of-the-art cosmological
simulations resolve subhaloes with force softenings that roughly
follow the scaling reflected by the yellow-shaded band. The reason is
that many simulations adopt the scaling relation of
\cite{Power.etal.03} to pick their softening length. According to that
relation, $\varepsilon \propto \Np^{-0.5}$, which is very similar to the
scaling of Eq.~(\ref{softscale}).  In other words, simulations of
different mass resolution typically scale their softening parameter
such that subhaloes of a given physical mass are resolved with a $\Np$
and $\varepsilon$ that fall (roughly) along the yellow-shaded band.
Along this band, though, results look fairly similar, indicating that
convergence is extremely slow, This implies that simply improving the
mass and force resolution of numerical simulations, following the
\cite{Power.etal.03} scaling relations, may give results that {\it
  appear} converged, but that in reality continue to suffer from
numerical artifacts.  As we discuss in \S\ref{sec:concl} below, this may
have profound implications for numerous areas of astrophysics.
\begin{figure*}
\includegraphics[width=0.95\hdsize]{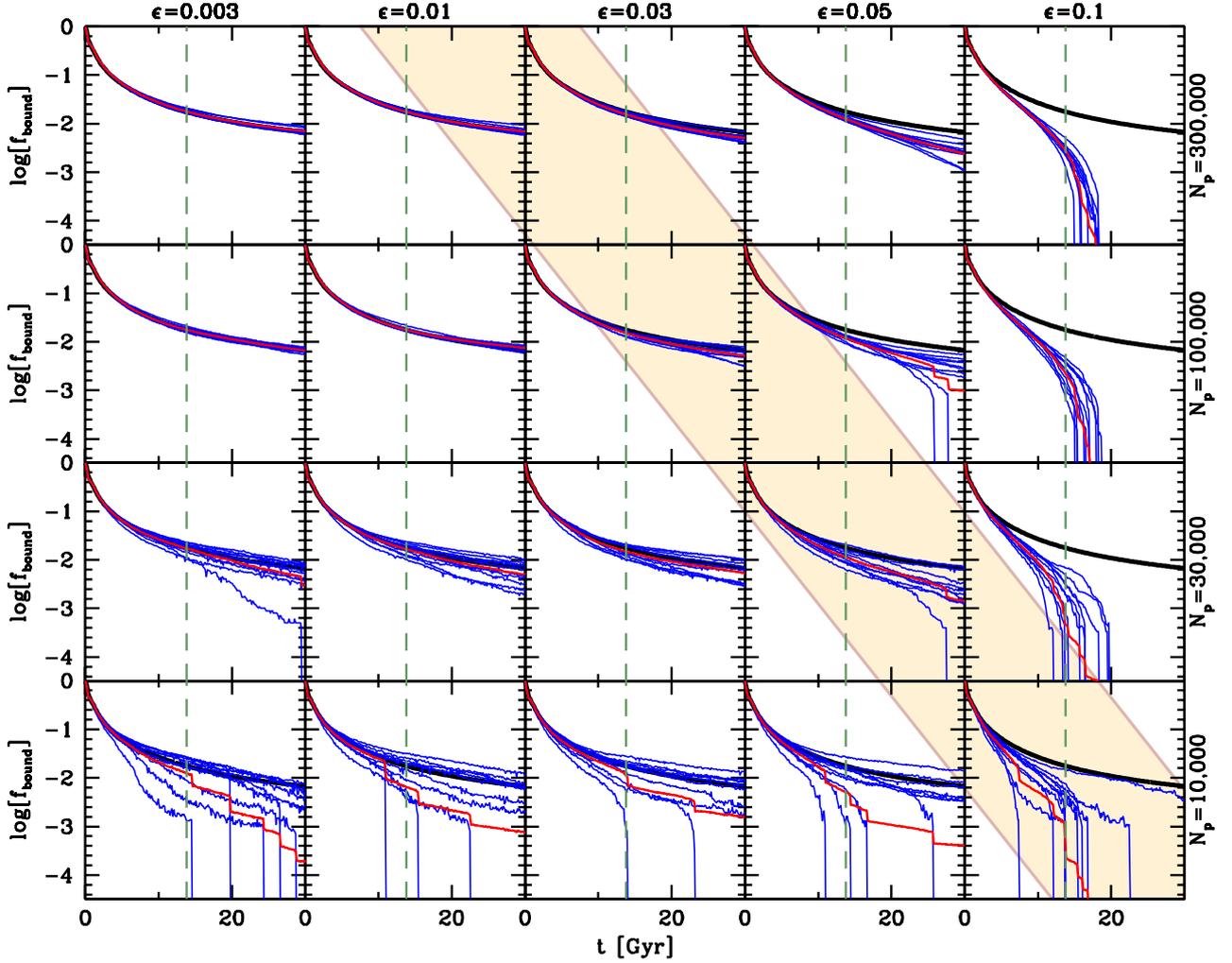}
\caption{Same as Fig.~\ref{fig:r01}, but for a circular orbit with
  $r_{\rm orb}/r_{\rm vir,h} = 0.2$.}
\label{fig:r02}
\end{figure*}

Fig.~\ref{fig:r02} shows the same results, but now for a circular
orbit with $\rorbNorm = 0.2$. As in Fig.~\ref{fig:r01}, the
yellow-shaded band corresponds to the scaling relation of
Eq.~(\ref{softscale}) and therefore indicates the typical
$(\Np,\varepsilon)$ with which haloes are resolved in typical
state-of-the-art cosmological simulations. Clearly, the requirements
to properly resolve the dynamics along this orbit are much weaker than
along an orbit with $\rorbNorm=0.1$: typically the dynamics are
properly converged for the duration of a Hubble time as long as $\Np
\gta 10^5$ and the softening length roughly follows the scaling of
\cite{Power.etal.03}. In fact, as long as $\Np \gta 10^5$ there is quite
a large range in softening values for which the results are
indistinguishable. Note also, though, that for $\Np<10^4$ the subhaloes
experience artificial disruption and are subject to severe
discreteness noise.
\begin{figure*}
\includegraphics[width=0.95\hdsize]{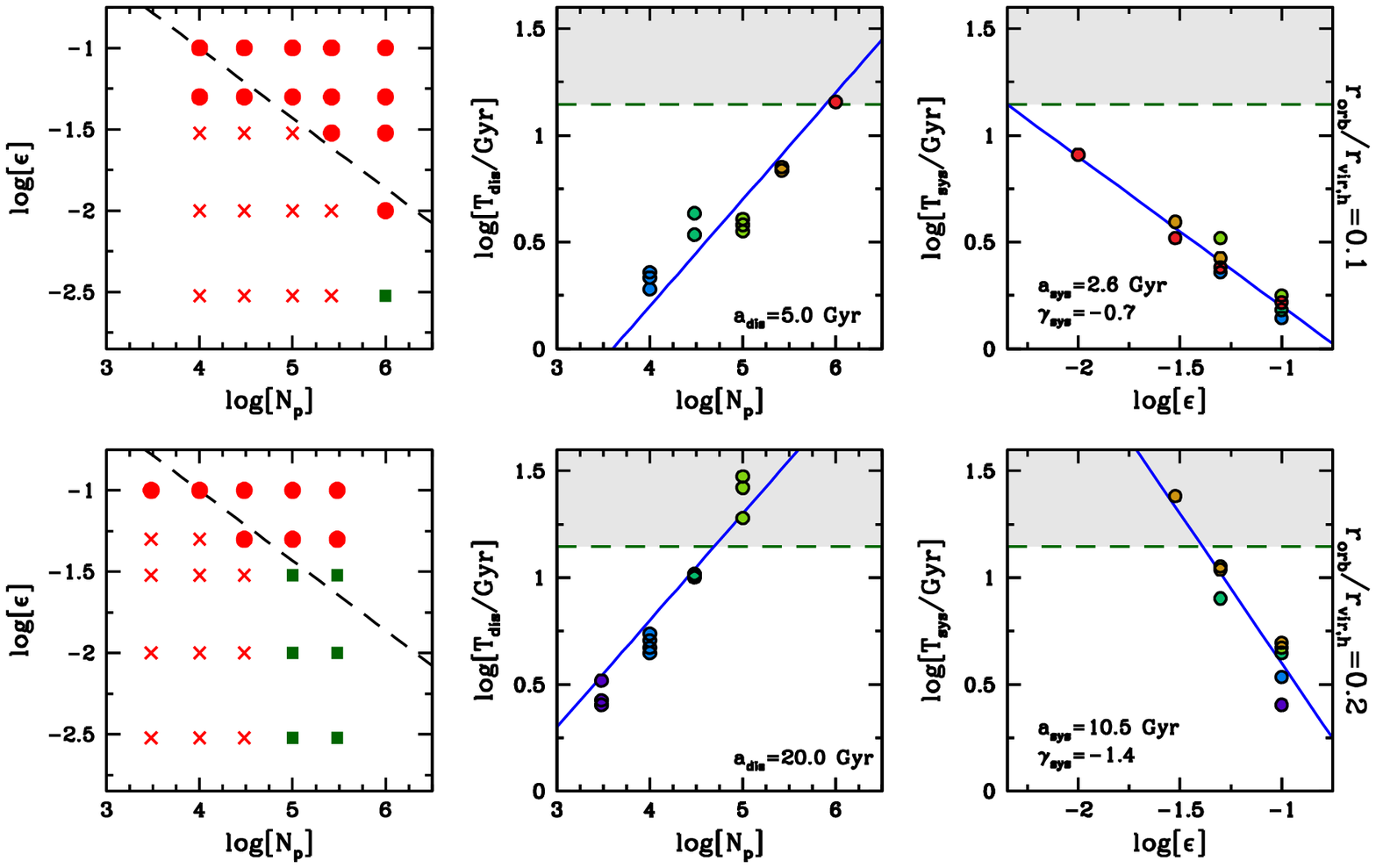}
\caption{{\it Left-hand panels:} For each set of 10 simulations in
  $(\Np,\varepsilon)$-parameter space, the symbols indicate whether
  the simulations suffer from discreteness noise (red crosses), from
  systematic errors (solid red circles), or whether the simulations
  are `converged' (solid green square). The dashed lines indicate the
  typical resolution with which dark matter (sub)haloes are resolved
  in state-of-the-art cosmological simulations
  (Eq.~[\ref{softscale}]). {\it Middle panels:} The discreteness time,
  $T_{\rm dis}$ as function of $\Np$, for those simulations with
  $T_{\rm dis} < T_{\rm sys}$. The grey-shaded area corresponds to
  time-scales longer than the Hubble time, while blue lines are fits
  to the data of the form $T_{\rm dis} = a_{\rm dis} N_5^{1/2}$, with
  the best-fit value of $a_{\rm dis}$ indicated in each panel.  {\it
    Right-hand panels:} The systematic time, $T_{\rm sys}$ as function
  $\varepsilon$ for those simulations with $T_{\rm sys} < T_{\rm
    dis}$.  Blue lines are fits of the form $T_{\rm sys} = a_{\rm sys}
  (\varepsilon/0.05)^{\gamma_{\rm sys}}$, with the best-fit values of
  $a_{\rm sys}$ and $\gamma_{\rm sys}$ as indicated.  In each column,
  top and bottom panels correspond to simulations with $\rorbNorm =
  0.1$ and $0.2$, respectively. In the middle and right-hand panels, the
  color of the symbols indicates $N_{\rm p}$.}
\label{fig:summ}
\end{figure*}

\subsection{Timescales}
\label{sec:timescales}

To make the results in Figs.~\ref{fig:r01} and~\ref{fig:r02} more
quantitative, we compute, for each $(\Np,\varepsilon,\rorbNorm)$ two
time scales. The first is the `systematic' time, $T_{\rm sys}$, which
we define as the time it takes for the average of the 10 random
realizations (the red curves in Figs.~\ref{fig:r01} and~\ref{fig:r02}) 
to deviate more than 0.1dex from the converged results (the thick black 
curves). Note that this deviation is
always in the sense of too much mass loss, which in many cases results
in artificial disruption. In addition, we also define the discreteness
time, $T_{\rm dis}$, as the time it takes for the standard deviation
among the 10 random realizations to become 0.1dex. The results are
shown in Fig.~\ref{fig:summ}, where the upper and lower panels
corresponds to $\rorbNorm = 0.1$ and $0.2$, respectively. The left
column shows for each $(\Np, \varepsilon)$ whether simulations with
these parameters suffer from systematic errors ($T_{\rm sys} < {\rm
  MIN}[T_{\rm dis}, t_\rmH]$; solid red circle), from discreteness
noise ($T_{\rm dis} < {\rm MIN}[T_{\rm sys}, t_\rmH]$; red cross), or
whether the simulations are `converged' ($t_\rmH < {\rm MIN}[T_{\rm
    sys},T_{\rm dis}]$; solid green square). The dashed lines
correspond to Eq.~(\ref{softscale}), and indicates the typical
resolution with which dark matter (sub)haloes are resolved in
state-of-the-art cosmological simulations. Note that for
$\rorbNorm=0.1$ these cosmological simulations are never converged and
typically suffer from systematic errors (too much mass loss and
artificial disruption). For $\rorbNorm=0.2$ the cosmological
simulations are converged as long as $\Np \gta 10^5$.

For those combinations of $\Np$ and $\varepsilon$ for which $T_{\rm
  dis} < T_{\rm sys}$ (red crosses and green squares in left-hand
panel), the middle panels of Fig.~\ref{fig:summ} plot $T_{\rm dis}$ as
function of $\Np$. The grey-shaded area corresponds to time-scales
longer than the Hubble time, while the blue lines are fits to the data
points of the form $T_{\rm dis} = a_{\rm dis} (\Np/10^5)^{1/2}$, with the
best-fit value of $a_{\rm dis}$ indicated in each panel. This
functional form is motivated by the notion that Poisson noise scales
with $\sqrt{\Np}$, and provides a reasonable fit to the data, thus
providing additional support for the notion that the run-away
instability is driven by discreteness noise. Note that $a_{\rm dis}$
depends strongly on $\rorbNorm$ and thus the strength of the tidal
field; the stronger the tidal field, the stronger the amplification of
the discreteness noise.

Finally, for those combinations of $\Np$ and $\varepsilon$ that result
in a systematic overestimation of the mass loss rate (red solid dots
in left-hand panels), the right-hand panels of Fig.~\ref{fig:summ} plot
$T_{\rm sys}$ as function of $\varepsilon$. Clearly $T_{\rm sys}$
depends strongly on the softening length used, and only very weakly on
$\Np$. The blue lines are fits of the form $T_{\rm sys} =
a_{\rm sys} (\varepsilon/0.05)^{\gamma_{\rm sys}}$, with the best-fit
values of $a_{\rm sys}$ and $\gamma_{\rm sys}$ as indicated. Clearly,
$T_{\rm sys}$ is much shorter in the presence of a stronger tidal
field, albeit with a much weaker dependence on $\varepsilon$.

\section{Criteria for Numerical Convergence}
\label{sec:criteria}

In the previous sections, we have shown how inadequate force softening
and discreteness noise can cause excessive mass loss and premature
disruption of dark matter substructure in numerical $N$-body
simulations. In this section we discuss under what conditions one may
deem individual subhaloes, in large-scale, cosmological simulations of
structure formation, adequately resolved. When analyzing such
simulations, these criteria can be used to discard subhaloes that are
affected by one or more of the numerical artefacts identified here,
and whose properties are therefore unreliable.

\subsection{Force Softening}
\label{sec:eps}

In \S\ref{sec:convergence} we defined $T_{\rm sys}$ as the time when
the {\it median} $\fbound$, obtained from a set of 10 simulations,
deviates 0.1 dex from the converged results. $T_{\rm sys}$ is
extremely sensitive to the force softening, and we therefore seek a
relation between the Plummer softening length, $\varepsilon$, and
properties of the subhalo in question that can be used to identify
$T_{\rm sys}$.
\begin{figure*}
\includegraphics[width=0.96\hdsize]{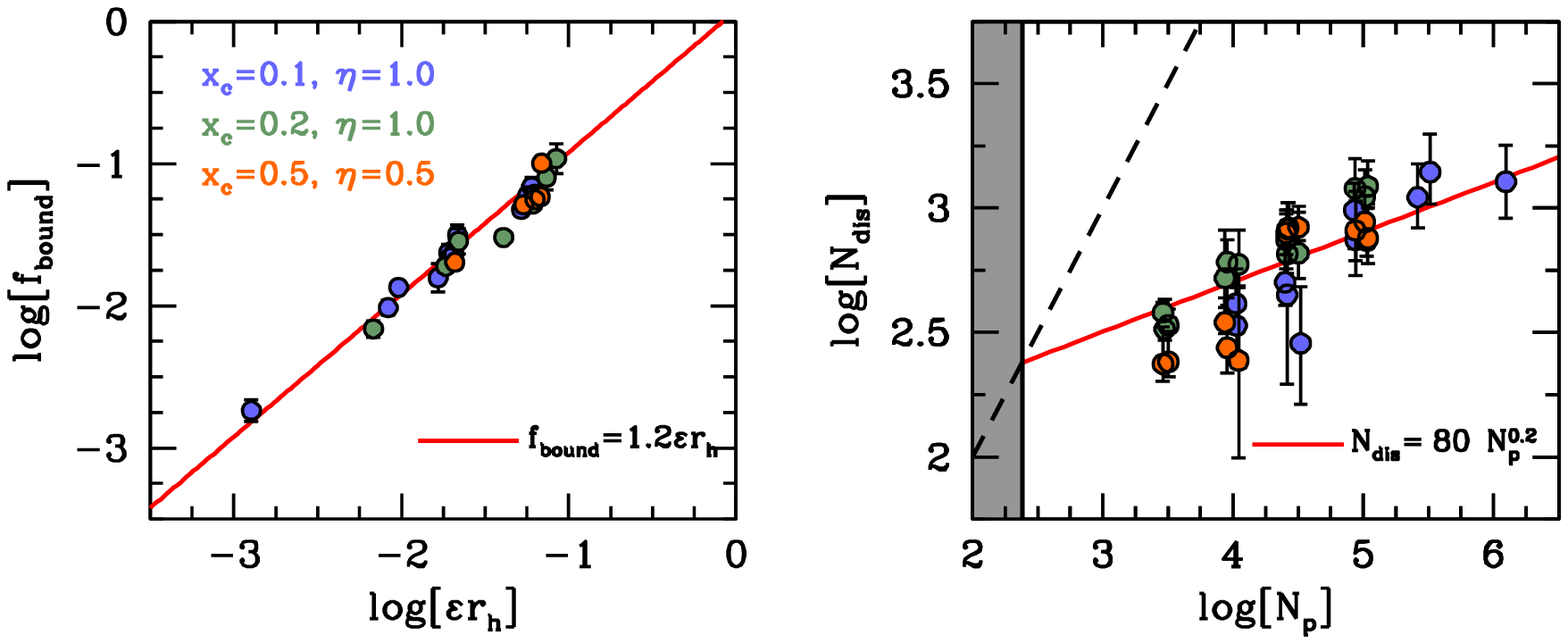}
\caption{{\it Left-hand panel:} The bound fraction at $T_{\rm sys}$ as
  a function of the product of the half-mass radius at that time,
  $r_\rmh$, and the softening length, $\varepsilon$. As is evident,
  these parameters are tightly correlated, indicating that $N$-body
  simulations fail to resolve the tidal evolution of subhaloes once
  they violate the conditions specified by criterion~(\ref{acccrit}).
  {\it Right-hand panel:} The average number of bound particles,
  $\Ndis$, in a subhalo at time $T_{\rm dis}$, when discreteness noise
  triggers a runaway instability.  Results are plotted as function of
  $\Np$, the number of particles in the initial subhalo. Errorbars
  indicate the 68 percent confidence intervals, and the solid, red
  line corresponds to the best-fit relation following the scaling
  expected for Poisson noise (see Appendix~\ref{App:dis}).  The dashed
  line corresponds to $\Ndis = \Np$, while the gray-shaded region
  indicates where $\Np<240$. Such subhaloes are adversely affected by
  discreteness noise from the moment they are accreted by their hosts.
  In both panels blue and green symbols correspond to simulations of
  circular orbits with $\rorbNorm = 0.1$ and $0.2$, respectively, while
  the orange symbols correspond to an eccentric orbit with $x_\rmc=\eta=0.5$.}
\label{fig:crit}
\end{figure*}

We are motivated by the work of \cite{Power.etal.03}, who introduced
the concept of a {\it characteristic} acceleration, $\acrit = G M_{\rm
  vir} / (\varepsilon r_{\rm vir}) = V^2_{\rm vir} / \varepsilon$. In
particular, they showed that NFW haloes {\it in isolation} are
resolved down to a convergence radius where $a(r) = G M(r) / r^2
\simeq \acrit/2$. At smaller radii $a(r) > \acrit/2$, and the mass
profile of the simulated halo is unreliable.

The acceleration profile, $a(r)$, of an NFW halo increases
monotonically inwards, asymptoting to a maximum value of
\begin{equation}\label{a0NFW}
\amax = \lim_{r \downarrow 0} \frac{G M(r)}{r^2} = 
\frac{c^2}{2 f(c)} \frac{G M_{\rm vir}}{r_{\rm vir}^2}
\end{equation}
where $c$ is the NFW concentration parameter, and
\begin{equation}\label{fx}
  f(x) = \ln(1+x) - \frac{x}{1+x},.
\end{equation}
We suspect that a subhalo will no longer be properly resolved once its
maximum acceleration, $\amax$, becomes smaller than of order the
characteristic acceleration, $\acrit$. Hence, we suspect that $T_{\rm
  sys}$ marks the time when the ratio $\acrit/\amax$ drops below some
critical value of order unity.

In the case of a subhalo, both $\acrit$ and $\amax$ evolve with time.
Let us first focus on $\acrit$. The original definition by
\cite{Power.etal.03} is based on the virial properties of the halo.
Since the virial properties of a subhalo undergoing tidal mass
stripping are ill-defined, we instead define the characteristic
acceleration based on the subhalo's half-mass radius, $r_\rmh$,
according to
\begin{equation}\label{achar}
\acrit \equiv \frac{G \, m_\rms(r_\rmh)}{\varepsilon \, r_\rmh} =
\frac{G \, \fbound \, \ms}{2 \, \varepsilon \, r_\rmh}\,.
\end{equation}
Here $\ms$ is the mass of the initial subhalo, i.e., prior to
being introduced to the tidal field. In the case of a cosmological
simulation, this roughly coincides with the mass of the subhalo at the
time of accretion into the host halo. Computing $\amax(t)$ for an
evolving subhalo in a numerical simulation requires taking the limit
of $r$ going to zero of the enclosed mass profile. Because of the
finite number of particles, computing this limit numerically is
extremely noisy.
Therefore instead we simply compute $\amax$ for the {\it initial}
subhalo using the analytical expression~(\ref{a0NFW}) for an NFW halo.
Note that the central density distribution of a subhalo only undergoes
relatively weak changes (cf. Fig.~\ref{fig:prof06}), and we therefore
expect this initial $\amax$ to be a reasonable representation of the
maximum, central acceleration of a subhalo at any time. As we show
below, this choice is further justified by the fact that indeed
$\acrit/\amax$ defined this way takes on a fixed value at $T_{\rm
  sys}$.

Based on the above considerations, we thus expect that $T_{\rm sys}$
corresponds to the time when
\begin{equation}\label{accrat}
  \frac{\acrit}{\amax} = \frac{\fbound f(c)}{(\varepsilon/r_{{\rms},0}) \,
    (r_\rmh/r_{{\rms},0})} = \chi_{\rm crit}\,.
\end{equation}
Here $r_{{\rms},0}$ is the scale-radius of the {\it initial} (i.e., at
accretion) subhalo, and the value of the free parameter, $\chi_{\rm
  crit}$, is to be determined from our numerical experiments. We do so
as follows: for each combination $(\Np,\varepsilon)$ for which the
simulations discussed in \S\ref{sec:convergence} suffer from
systematic errors (i.e., $T_{\rm sys} < T_{\rm dis}$), we compute 
$r_\rmh$ and $\fbound$ at the time
$T_{\rm sys}$.  The left-hand panel of Fig.~\ref{fig:crit} plots the
resulting $\log\fbound$ versus $\log[\varepsilon \, r_\rmh]$ (in model
units, for which $r_{{\rms},0}=1$). Errorbars on $r_\rmh$ and
$\fbound$ are obtained from the scatter in these properties among the
10 simulations at $T_{\rm sys}$, and are typically smaller than the
symbols. Clearly, there is an extremely tight relation between these
parameters, which is independent of the subhalo's orbit and well
described by
\begin{equation}
  \fbound = 1.2 \, \left(\frac{\varepsilon}{\rs}\right) \, 
  \left(\frac{r_\rmh}{\rs}\right)\,,
\end{equation}
shown as the solid, red line.  Substitution in Eq.~(\ref{accrat}), and
using that the subhaloes in our simulation all have an initial NFW
concentration parameter $c=10$, we infer that $\chi_{\rm crit} \simeq
1.79$. We thus conclude that dark matter subhaloes in numerical
$N$-body simulations are only properly resolved as long as
\begin{equation}\label{acccrit}
\fbound > \frac{1.79}{f(c)} \, 
\left(\frac{\varepsilon}{\rs}\right) \,
\left(\frac{r_\rmh}{\rs}\right)\,
\end{equation}
where $r_{{\rms},0}$ and $c$ are the scale radius and concentration
parameter of the (NFW) subhalo {\it at accretion}.

We have experimented with a number of alternative criteria, including
ones in which we attempt to determine the instantaneous value of
$\amax$ from the numerical subhaloes. However, none of these
alternatives faired any better than criterion~(\ref{acccrit}).  The
only alternative that performed almost equally well is a simple
criterion based on the ratio $r_\rmh/\varepsilon$. Using the same set
of simulations as above we find that at $T_{\rm sys}$,
$\log(r_\rmh/\varepsilon) = 0.84 \pm 0.14$. This suggests that
subhaloes are resolved as long as $r_\rmh/\varepsilon >
6.9^{+2.6}_{-1.9}$. As it turns out, this is simply another way of
writing criterion~(\ref{acccrit}). The reason is that our simulations
reveal a reasonably tight relation between $\fbound$ and
$r_\rmh/\rh$ which is well approximated by 
\begin{equation}\label{rhfb}
\frac{r_\rmh}{\rh} \simeq 0.7 \, \fbound^{0.5}\,
\end{equation}
over the range where $-2.5 \lta \log(\fbound) \lta -0.5$.
Substitution in Eq.~(\ref{acccrit}) and using the relation between the
half-mass radius and scale-radius of an NFW halo (Eq.~[\ref{rhrs}]),
we can rewrite Eq.~(\ref{acccrit}) as
\begin{equation}\label{rhcrit}
  \frac{r_\rmh}{\varepsilon} > 0.62 \, \frac{c^{1.26}}{f(c)}\,.
\end{equation}
For $c=10$, the value adopted for all subhaloes in our simulations,
the right-hand side of this criterion is equal to 7.6, in excellent
agreement with the average ratio of $r_\rmh/\varepsilon$ inferred from
our simulations at $T_{\rm sys}$. Since relation~(\ref{rhfb}) reveals
a fair amount of variance among simulations, especially for small
values of $\fbound$, we generally recommend using
criterion~(\ref{acccrit}) over criterion~(\ref{rhcrit}).

\subsection{Discreteness Noise}
\label{sec:Ndis}

We now derive a criterion that can be used to check whether a subhalo
in a numerical $N$-body simulation is significantly affected by
discreteness noise.  Since this is purely a manifestation of finite
particle number, we seek a criterion that depends on the number of
particles that is used to model the subhalo. In
\S\ref{sec:convergence} we defined the characteristic time $T_{\rm
  dis}$ as the time when the standard deviation in $\fbound$ among 10
simulations becomes 0.1 dex. We now define $N_{\rm dis}$ as the
average number of particles in the subhalo remnant at this time
$T_{\rm dis}$.

We compute $N_{\rm dis}$ for each combination of $(\Np,\varepsilon)$
for which the simulations suffer from discreteness noise (i.e.,
$T_{\rm dis} < T_{\rm sys}$). Since
$\sigma_{\log f}$ is estimated from only 10 simulations, it carries an
error, which in turn imposes an error on $N_{\rm dis}$. To estimate
this error we proceed as follows. Under the assumption that
$\log\fbound$ at any given epoch follows a log-normal distribution,
the fractional error on the standard deviation estimated from $N$
realizations of that distribution is equal to $1/\sqrt{2(N-1)}$
\citep[e.g.,][]{Taylor.97}. Hence, for $N=10$ we have that the
fractional error on $\sigma_{\log f}$ is 0.236.  We then estimate the
upper and lower bounds of the 68 percent confidence interval on
$N_{\rm dis}$ as the values for $\langle N_{\rm dis} \rangle$ at the
times when $\sigma_{\log f} = 0.076$ and $\sigma_{\log f} = 0.124$,
respectively.
 
The results are shown in the right-hand panel of Fig.~\ref{fig:crit},
which plots $\Ndis$ (with errorbars indicating the 68 percent
confidence intervals) as a function of $\Np$, the number of particles
in the {\it initial} subhalo. Typically $300 \lta \Ndis \lta 1000$, with a
weak trend of increasing $\Ndis$ with increasing $\Np$. As discussed
in Appendix~\ref{App:dis}, because of Poisson noise in the mass loss
rate of $N$-body subhaloes, one expects that $\Ndis \propto
\Np^{0.2}$. Fitting such a relation to the data, we obtain $\Ndis = 80
\, \Np^{0.2}$, which is indicated by the red line.

Hence, in a cosmological simulation one can guard against subhaloes
that are potentially affected by discreteness noise by only selecting
subhaloes that obey the following criterion:
\begin{equation}\label{disccrit}
N > 80 \, N_{\rm acc}^{0.2}\,.
\end{equation}
Here $N_{\rm acc}$ is the number of particles in the subhalo at its
moment of accretion. Note that subhaloes with
$N_{\rm acc} < 80^{5/4} \simeq 240$ always violate this criterion, and
therefore are susceptible to discreteness noise immediately after
accretion. In other words, $N$-body simulations are unable to properly
resolve the tidal evolution of dark matter subhaloes with $N \lta 250$
particles at accretion.
\begin{figure}
\includegraphics[width=\hssize]{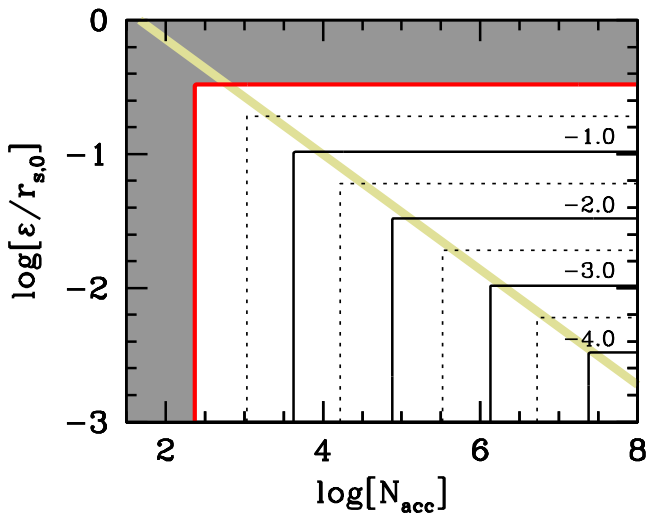}
\caption{Contour plot of the minimal bound mass fraction,
  $\fbound^{\rm min}$, defined by Eq.(\ref{fboundmin}), in the
  parameter space of $N_{\rm acc}$, the number of particles of the
  subhalo at accretion, and $\varepsilon/\rs$, the softening length in
  units of the subhalo's scale radius at accretion. Subhaloes with an
  instantaneous $\fbound < \fbound^{\rm min}$ are no longer reliable,
  and should be discarded from any analysis. Contours correspond to
  $\log\fbound^{\rm min} = (-4, -3.5, -3,...,0)$, with solid contours
  corresponding to integer values, as indicated.  For comparison, the
  yellow-shaded band corresponds to Eq.~(\ref{softscale}), and
  indicates the softening length with which subhaloes of given $N_{\rm
    acc} = N_{200}$ are resolved in the Millennium simulation.}
\label{fig:crit_cont}
\end{figure}

\subsection{Eccentric Orbits}
\label{sec:ecc}

Thus far we have focused exclusively on circular orbits.
In reality, though, subhaloes move on eccentric orbits with fairly
large apo-to-pericentre ratios \citep[e.g.][]{Tormen.97, Ghigna.etal.98,
vdBosch.etal.99}. As a consequence, they are significantly impacted by 
impulsive heating due to tidal shocks at pericentric passage (see Paper~I), 
something that is not accounted for in our simulations of 
circular orbits.  This begs the question to what extent the criteria derived 
above are applicable to the more realistic, eccentric orbits. In order 
to address this, we have performed a set of simulations, similar to those 
used above, but for an eccentric orbit with $x_\rmc \equiv r_\rmc(E)/\rvir 
= 0.5$ and an orbital circularity of $\eta \equiv L/L_\rmc(E) = 0.5$. Here 
$E$ and $L$ are the orbital energy and angular momentum, respectively, 
$r_\rmc(E)$ is the radius of a circular orbit of energy $E$, and 
$L_\rmc(E)$ is the corresponding angular momentum.  Note that a circularity 
of $\eta=0.5$ is characteristic of orbits at infall 
\citep[e.g.][]{Zentner.etal.05, Wetzel.11, Jiang.etal.15, vdBosch.17},

We adopt the fiducial parameters for the masses and concentrations of 
the host and subhalo (see Table~1), and run sets of ten simulations each 
for different $\Np$ and $\varepsilon$. For each of these sets, we 
determine $T_{\rm sys}$ and $T_{\rm dis}$ as described in 
\S\ref{sec:timescales}, and we test whether these simulations are subject 
to the same criteria as the circular orbits discussed above. 
The results, based on a sample of 200 simulations with
$3000 \leq \Np \leq 100,000$ and $0.003 \leq \varepsilon \leq 0.1$ are
shown as the orange symbols in Fig.~\ref{fig:crit}. Note how the
results for the eccentric orbit are in excellent agreement with 
the relations for the circular orbits. This strongly suggests that 
criteria~(\ref{acccrit}) and~(\ref{disccrit}) are valid for orbits of
all energies and angular momenta (see \S\ref{sec:caveats}).

\subsection{The Minimal Bound Mass Fraction}
\label{sec:fmin}

Both criteria~(\ref{acccrit}) and~(\ref{disccrit}) can be cast in a
constraint on the bound mass fraction, $\fbound$, for a given set
of numerical parameters $(\Np,\varepsilon)$. For 
criterion~(\ref{acccrit}) this is achieved by using Eqs.~(\ref{rhfb})
and~(\ref{rhrs}), which yields
\begin{equation}\label{fboundminA}
  \fbound > \fbound^{\rm min,1} = 1.12 \, \frac{c^{1.26}}{f^2(c)} \,
  \left(\frac{\varepsilon}{\rs}\right)^2\,.
\end{equation}
For criterion~(\ref{disccrit}) one simply divides both sides by $N_{\rm acc}$
to obtain
\begin{equation}\label{fboundminB}
  \fbound > \fbound^{\rm min,2} = 0.32 \, 
  \left(\frac{N_{\rm acc}}{1000}\right)^{-0.8}\,.
\end{equation}
We now define
\begin{equation}\label{fboundmin}
  \fbound^{\rm min} \equiv {\rm MAX}[\fbound^{\rm min,1},\fbound^{\rm min,2}]\,
\end{equation}
such that subhaloes with an instantaneous bound mass fraction $\fbound
> \fbound^{\rm min}$ may be deemed converged, while those subhaloes
for which $\fbound$ is smaller than this minimal value are unreliable;
either because of discreteness noise, or because of inadequate force
softening.

Fig.~\ref{fig:crit_cont} shows contours of $\fbound^{\rm min}$ in the
$(N_{\rm acc},\varepsilon)$ parameter space. This figure can be used
to read off whether a particular subhalo in a cosmological simulation
may be deemed converged or not. Note that the dark-grey shaded region
indicates the part of parameter space for which $\fbound^{\rm min} >
1$; subhaloes with these parameters can never be deemed converged 
(i.e., $N_{\rm acc} \lta 250$ or $\varepsilon \gta 0.3 \rs$).
For comparison, the yellow-shaded band corresponds to
Eq.~(\ref{softscale}), and indicates the softening length with which
subhaloes of given $N_{\rm acc} = N_{200}$ are resolved in the
Millennium simulation. It shows that subhaloes in this simulation with
$\fbound < 0.01$ ($<0.001$) can only be deemed converged if $N_{\rm
  acc} \gta 10^5$ ($>10^6$). 

Finally, we emphasize that although the results in
Fig.~\ref{fig:crit_cont} depend on the concentration parameter of the
subhalo at accretion (here we have adopted $c=10$), this dependence is
very weak: for $5 < c < 30$ (the typical range of concentration
parameters for CDM haloes), the factor $c^{1.26}/f^2(c)$ only ranges
between 8 and 12, and therefore has almost negligible impact on the
value of $\fbound^{\rm min}$.

\section{Sensitivity to Initial Conditions}
\label{sec:ICsens}

In \S\ref{sec:num} we discussed how the numerical parameters $\Delta
t$, $\theta$, $\varepsilon$ and $\Np$, impact the outcome of numerical
$N$-body simulations of the tidal evolution of dark matter
substructure. In addition to setting these numerical parameters, the
simulator also needs to decide on how to pick and set the initial
conditions (ICs) for the simulation. As already eluded to in
\S\ref{sec:ICs}, this is non-trivial. In this section we therefore
examine how the simulation outcome depends on subtle issues related to
setting the ICs. We start with the actual method that is used to pick
the phase-space coordinates of the particles that make up the
subhalo. Next we discuss the impact of letting the subhalo relax prior
to placing it in the external tidal field of the host halo, and the
impact of truncating the subhalo at its tidal radius.
\begin{figure*}
\includegraphics[width=0.96\hdsize]{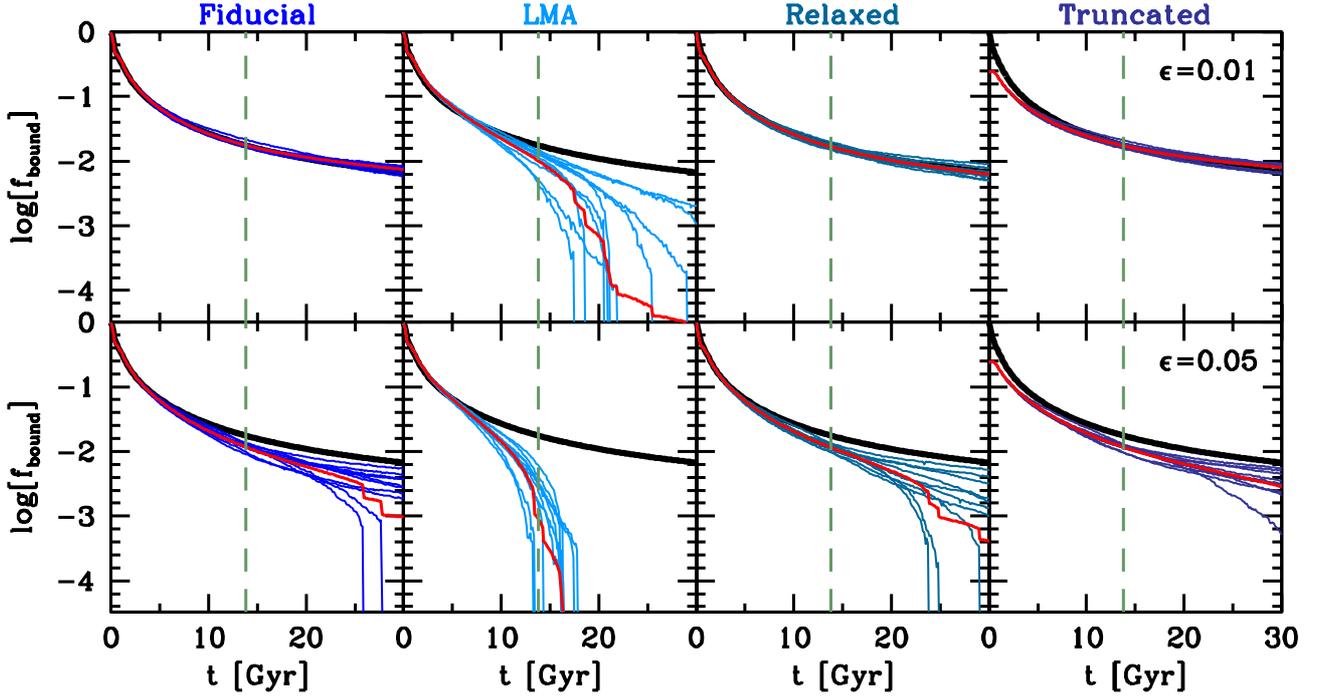}
\caption{Blue lines show the bound mass fractions as function of time
  for sets of 10 simulations with $\Np=10^5$ each, of subhaloes on a
  circular orbit with $\rorbNorm = 0.2$ in a host halo with $M_\rmh =
  1000 \ms$ and $c_\rmh = 5.0$. The red line shows the corresponding
  average, while the solid, black line shows converged results
  from a simulation with $\Np=10^7$ and $\varepsilon = 0.003$. Top and
  bottom panels correspond to simulations run with a force softening
  length of $\varepsilon = 0.01$ and $0.05$, respectively. Different
  columns correspond to different methods used to initialize the
  subhaloes. The columns labeled `Fiducial' and `LMA' shows results
  in which the initial positions and velocities are drawn from the
  distribution function, $f(E)$, and based on the local Maxwellian
  approximation, respectively. In the case of the columns labeled
  `Relaxed' and `Truncated' the initial subhaloes were allowed to
  relax for 10 Gyr and truncated at the tidal radius, respectively,
  prior to being placed in the tidal field of the host halo. See text
  for details.}
\label{fig:IC}
\end{figure*}

All simulations presented below use the fiducial physical and
numerical parameters of Table~1. In addition, for comparison we also
run each of these simulations with a five times smaller softening
length, $\varepsilon = 0.01$.

\subsection{Local Maxwellian Approximation}
\label{sec:LMA}

In all simulations discussed thus far we have set-up the initial
conditions for the NFW halo by sampling the isotropic distribution
function, $f(E)$, as described in \S\ref{sec:ICs}. An alternative
method, which has been applied in numerous studies
\citep[e.g.,][]{Hernquist.93, Springel.White.99, Klypin.etal.99a,
  Mayer.etal.01, Penarrubia.etal.02, Taffoni.etal.03,
  Hayashi.etal.03}, is to use the local Maxwellian approximation
(hereafter LMA), which is based on the ansatz that the velocity
distribution at any given position can be approximated by a
multivariate Gaussian, whose velocity dispersion follows from solving
the Jeans equation at that position. In detail, one proceeds as
follows. One first draws a radius, $r$, using the enclosed mass
profile of the halo in question, followed by two angles that allow one
to assign the particle Cartesian coordinates. Next, one computes the
local, radial velocity dispersion, $\sigma^2_r(r)$, by solving the
Jeans equation \citep[for an NFW halo, the solution is given by
  Eq.~11 in][]{vdBosch.etal.04}. Assuming isotropy, one then obtains
the local Cartesian velocities by drawing three Gaussian deviates with
a dispersion equal to $\sigma_r(r)$. We follow \cite{Hernquist.93},
and redraw the Cartesian velocity components if the resulting 3D speed
exceeds 95 percent of the local escape speed. The advantage of this
method over our fiducial one is that it is easier to implement and
avoids having to compute and sample the DF \citep[see discussion
  in][]{Hernquist.93}. However, the method is only approximate, in
that it ignores higher order moments of the collisionless Boltzmann
equation, and it has been demonstrated that the local Maxwellian
approximation can yield ICs that are far from equilibrium; as a
consequence, N-body realizations of such models rapidly relax to a
steady-state that can differ significantly from the initial, intended
one \citep[][]{Kazantzidis.etal.04}.

Most importantly, \cite{Kazantzidis.etal.04} compared the tidal
evolution of two subhaloes on the same orbit; one whose ICs were
obtained using the LMA, and the other for which the phase-space
coordinates are drawn self-consistently from the (isotropic)
distribution function, $f(E)$.  They find that the subhalo constructed
under the assumption that the local velocity field is Maxwellian is
completely disrupted in a few orbits, whereas the self-consistent
subhalo survives for the entire duration of the simulation.
\cite{Kazantzidis.etal.04} therefore conclude that the LMA method for
generating ICs results in a systematic bias regarding the tidal
evolution of substructure. However, there is an important caveat here,
in that they simulated each of their subhaloes with only $\Np=10^5$
particles. As demonstrated in this paper, such simulations are subject
to severe discreteness noise. Hence, the difference in $\fbound(t)$
for the two subhaloes might simply reflect discreteness noise, i.e.,
the LMA-initialized subhalo, {\it by chance}, happened to undergo
disruption due to the discreteness-driven run-away instability, rather
than due to a {\it systematic} difference arising from the different
methods used to draw the initial phase-space coordinates.

To test this, we construct two {\it ensembles} of 10 subhaloes with
$\Np=10^5$ each. The first is our `fiducial' ensemble , in which the
initial positions and velocities of the particles are drawn from the
DF, $f(E)$, using our fiducial method outlined in
\S\ref{sec:ICs}. Ensemble members only differ in their random
realizations. The second ensemble consists of random realizations
based on the LMA method described above. Note that both methods assume
that the distribution function if isotropic. For each of these 20
subhaloes we run two simulations; one with $\varepsilon = 0.01$ and
the other with $\varepsilon=0.05$. In each case the subhalo is placed
on a circular orbit with $\rorbNorm = 0.2$ in a host halo with $M_\rmh
= 1000 \ms$ and $c_\rmh = 5.0$, and evolved for $30$ Gyr.

The results are shown in Fig.~\ref{fig:IC} where the first and second
columns from the left show the results for the fiducial and LMA
ensembles, respectively. There is a clear difference, in that the
LMA-based subhaloes experience {\it systematically} more mass loss,
almost always resulting in premature disruption. This ratifies the
conclusion by \cite{Kazantzidis.etal.04}, that the local Maxwellian
approximation for setting up the initial conditions of subhaloes,
results in accelerated mass loss, leading to their artificial
disruption. As discussed in \cite{Kazantzidis.etal.04}, the reason for
this systematic difference is that subhaloes initialized using the LMA
method are far from equilibrium, and quickly relax to a state with a
much shallower central density distribution.

\subsection{Initial Relaxation}
\label{sec:relax}

As discussed in \S\ref{sec:ICs}, the initial haloes in our fiducial
ensemble are not in perfect equilibrium.  First of all, we
instantaneously truncate the halo at its own virial radius, without
accounting for this truncation in $f(E)$. Secondly, we do not account
for the simulation's force softening when setting up the initial
conditions. And thirdly, by placing the initial halo in a tidal field,
we instantaneously introduce tidal perturbations that are not
accounted for in the ICs.

To gauge the impact of these shortcomings, we perform two different
tests. First, we perform a set of simulations in which we first evolve
the subhaloes in our fiducial ensemble for 10 Gyr in isolation, prior
to placing them in the tidal field of their host haloes (using the
same $\rorbNorm=0.2$ orbit as above). This allows the subhalo to
relax, and to reach virial equilibrium, before being introduced to its
external, tidal field. The results, again for $\varepsilon = 0.01$ and
$0.05$, are shown in the third column of Fig.~\ref{fig:IC}. Note that
these are virtually indistinguishable from our fiducial simulations
(left-most column), indicating that our simulation results are not
influenced by the fact that their initial conditions are not in
perfect equilibrium.


\subsection{Initial Tidal Truncation}
\label{sec:trunc}

As a second test, we truncate the initial subhalo at its tidal radius,
which for a circular orbit at radius $r_{\rm orb}$ is given by
\begin{equation}\label{rt}
\rtid = r_{\rm orb} \, \left[ \frac{\ms(\rtid)/M_\rmh(r_{\rm orb})}{
     3 - \frac{\rmd \ln M_\rmh}{\rmd \ln r}\vert_{r_{\rm orb}}} \right]^{1/3} \,,
\end{equation}
(see Paper~I for details).  One might argue that such an initial
truncation is more realistic, as the subhalo is expected to have
already experienced tidal stripping on its way to the initial position
in the host halo. And if stripping were to simply remove all particles
beyond its tidal radius, this `tidal truncation' seems
appropriate. Indeed, some studies have used a similar `pre-truncation'
in their simulations \citep[e.g.,][]{Choi.etal.07}.

To test the impact of this `initial truncation' we proceed as follows.
For each subhalo in our fiducial ensemble, which has $\Np=10^5$ and is
initially truncated at its virial radius, we remove all particles with
$r > \rtid$. In the case of $\rorbNorm=0.2$ and our fiducial halo
concentrations (see Table~1), $\rtid = 0.183 r_{\rm vir,s}$, which
encloses on average about 26.5 percent of the subhalo's virial
mass. Hence, the average number of particles with which this new
ensemble of `tidally truncated' subhaloes is resolved is $\Np =
26,500$. The results of integrating each of these subhaloes for 30 Gyr
along a circular orbit with $\rorbNorm=0.2$ are shown in the
right-most column of Fig.~\ref{fig:IC}. Once again, the results are
very similar to those of our fiducial ensemble, although some
differences are apparent for the simulations with $\varepsilon =
0.05$. These arise from differences in the amount of dynamical
self-friction, which is suppressed in the tidally truncated ensemble
compared to the fiducial ensemble. The reason is that there is far
more tidally stripped material in the latter, resulting in a larger
reduction of the orbital radius. However, the effect is extremely
small, and does not have a significant impact on any of the simulation
results presented here.

To summarize, we confirm the conclusion of \cite{Kazantzidis.etal.04}
that using the local Maxwellian approximation to initialize subhaloes
results in excessive mass loss and premature, artificial disruption.
Although the ICs of our subhaloes are not in perfect equilibrium, we
have shown that this has no discernible impact on the final outcome of
our simulations. Also, the fact that the subhalo is instantaneously
introduced into the external tidal field of the host halo, does not
have a significant impact, in that pre-truncating the subhalo at its
tidal radius yields very similar results. We thus conclude that the
simulation results presented in this paper are meaningful and relevant
despite the non-physical nature of the ICs.

\section{Potential Caveats}
\label{sec:caveats}

The results presented here and in Paper~I suggest that most, if not
all, subhalo disruption in numerical simulation is numerical, and cast
some serious doubt on the belief that the subhalo mass functions in
cosmological numerical simulations are properly converged (i.e., are
reliable) down to a mass limit of 50-100 particles. However, before we
can claim that existing simulations are significantly in error, we
have to address a number of caveats:
\begin{itemize}

\item {\bf Parameter Space:} One of the main limitations of this study
  is that we have only covered a tiny fraction of parameter space. In
  particular, we focused (almost) exclusively on circular orbits, while
  keeping the masses and concentration parameters of the host and
  subhalo fixed to the fiducial values listed in Table~1.  In reality
  subhalo orbits span a wide range of energies and angular momenta.
  Typically, along more eccentric orbits the subhaloes experience more
  impulsive heating due to tidal shocks at pericentric passage\footnote{As 
    shown in Paper~I, the impulsive heating due to encounters with other 
    subhaloes, sometimes called `harassment' is negligible in comparison.}, 
   something that is absent along the circular orbits examined here. 
  The tidal evolution of a subhalo also strongly depends on the 
  concentrations of both the subhalo and the host halo, while the mass 
  ratio of the host and sub-halo mainly controls the strength of dynamical 
  friction (and self-friction). We emphasize, though, that our results 
  assess the impact of numerics relative to high-resolution, converged 
  simulations. Since we do not expect this {\it relative} behavior to be 
  strongly affected by these physical parameters, we conjecture that our 
  results remain valid for most of the relevant parameter space. Although 
  our test based on a single eccentric orbit (\S\ref{sec:ecc}) clearly 
  supports this conjecture, ideally this should be repeated for 
  other orbital configurations, mass ratios and/or halo concentration 
  parameters. However, given the limiting amount of resources and 
  man-power available, and given that the study presented here is already 
  based on many hundreds of simulations, such an extension is beyond the 
  scope of the research presented here. 
  
\item{\bf Host Halo Realism:} One aspect of our simulations that may
  potentially impact some of the outcome is the fact that we treated
  the host halo as a spherical, static and analytical potential. In
  reality, haloes are non-spherical, dynamical objects, typically
  growing in mass, undergoing mergers and violent relaxation. This
  impacts the orbits and hence the amount of tidal stripping and
  heating. In addition, in a `live' host halo (i.e., modeled as an
  $N$-body system, rather than an analytical potential), the subhalo
  will experience dynamical friction, in addition to the self-friction
  briefly discussed in \S\ref{sec:dynfric}. Although such dynamical
  friction is not significant for the mass ratio considered here
  ($M_\rmh/\ms = 1000$), it will play an important role for systems
  where $M_\rmh/\ms \lta 30$ \citep[e.g.,][]{MBW10}. Unfortunately,
  the only way to take proper account of all these effect is by
  simulating the system in its proper cosmological setting, and thus
  to run a cosmological simulation. However, doing so with the
  resolution required to achieve proper convergence, as specified
  here, is beyond the capabilities of present-day technology.

\item{\bf Impact of baryons:} Finally, we emphasize that this work has
  focused exclusively on dark matter only, without taking account of
  potential baryonic effects.  There is a rapidly expanding literature
  on how baryons may impact the abundance and demographics of dark
  matter substructure\citep[e.g.,][]{Maccio.etal.06, Weinberg.etal.08,
    Dolag.etal.09, Arraki.etal.14, Brooks.Zolotov.14,
    Despali.Vegetti.16, Fiacconi.etal.16, Wetzel.etal.16,
    Garrison-Kimmel.etal.17}.  Our work, in no way, aims to undermine
  the importance of baryons.  However, before we can make reliable
  predictions for how baryonic processes modify substructure in the
  dark sector, we first need to establish a better understanding of
  how tides impact substructure, and develop tools and criteria to
  assess the reliability of simulations (be it hydro or
  $N$-body). Hence, this work is to be considered a necessary first
  step towards a more reliable treatment of dark matter substructure
  (including satellite galaxies); not the final answer.

\end{itemize}

\section{Conclusions \& Discussion}
\label{sec:concl}

As part of our ongoing effort to understand the origin of the
ubiquitous disruption of dark matter subhaloes in numerical
simulations \citep[see][]{vdBosch.17}, we have performed a large suite
of hundreds of idealized simulations that follow the tidal evolution
of individual $N$-body subhaloes in a fixed, analytical host halo
potential.  The goal of these idealized, numerical experiments is not
to simulate realistic astrophysical systems, but rather to gain a
physical understanding of the complicated, non-linear and numerical
processes associated with the tidal stripping of dark matter
subhaloes. This is facilitated by restricting ourselves to purely
circular orbits (i.e., no tidal shock heating), and by focusing
exclusively on dark matter (i.e., no baryons).

By varying the orbital radius (strength of the tidal field), the
number of particles (mass resolution), the force softening (force
resolution), the tree opening angle (force accuracy) and the time step
used (time resolution), we address the following specific questions:
(i) is the disruption artificial (numerical) or real (physical),  (ii)
under what conditions do subhaloes disrupt, and what causes it, and
(iii) what are the numerical requirements to properly resolve the
tidal evolution of dark matter substructure?

We confirm the conclusions from our mainly analytical assessment in
Paper~I that most, if not all, disruption of substructure in $N$-body
simulations is numerical in origin \citep[see also][all of whome have
  argued that complete disruption of CDM subhaloes is extremely
  rare]{Kazantzidis.Moore.Mayer.04, Goerdt.etal.07, Diemand.etal.07a,
  Diemand.etal.07b, Penarrubia.etal.10}. As long as a subhalo is
resolved with sufficient mass and force resolution, a bound remnant
survives, even if the subhalo has lost more than 99.9 percent of its
original (infall) mass. We thus conclude that state-of-the-art
cosmological simulations still suffer from significant overmerging.
We have demonstrated that this is mainly due to inadequate
force-softening. In addition, we have shown that subhaloes in $N$-body
simulations are susceptible to a runaway instability which is
triggered by the amplification of discreteness noise in the presence
of a tidal field. These two processes conspire to put some serious
limitations on the reliability of dark matter substructure in
state-of-the-art cosmological simulations. In addition, in
cosmological simulations discreteness effects also inject artificial 
structure during the collapse of sheets and filaments 
\citep[e.g.,][]{Weinberg.93, Angulo.etal.13, Power.etal.16}, which 
further muddles the waters.

We have used our large suite of simulations to gauge under what
conditions inadequate force softening and discreteness noise impact
subhaloes. We find that properly resolving the tidal evolution of a
subhalo on a circular orbit at 10 percent of the host halo's virial
radius, requires that the subhalo is simulated with at least
$\Np=10^6$ particles and with a softening length, $\varepsilon$, that
is only $\sim 0.003$ times the subhalo's (NFW) scale radius,
$r_\rms$. These requirements, though, depend strongly on the strength
of the tidal field. For example, on a circular orbit at 20 percent of
the host halo's virial radius, $\Np=10^5$ and $\varepsilon/r_\rms =
0.03$ suffice to properly resolve its tidal evolution for at least a
Hubble time. More generally, we find that subhaloes in numerical
simulations start to be significantly affected by discreteness noise
once
\begin{equation}\label{disccritINV}
\fbound < 0.32 (N_{\rm acc}/1000)^{-0.8}\,,
\end{equation}
where $N_{\rm acc}$ is the number of particles in the subhalo at
accretion. Similarly, subhaloes in numerical simulations are
systematically affected by inadequate force resolution (often leading
to artificial disruption), once
\begin{equation}\label{acccritINV}
  \fbound < \frac{1.79}{f(c)} \, \left(\frac{\varepsilon}{\rs}\right) \,
\left(\frac{r_\rmh}{\rs}\right) \,,
\end{equation}
where $c$ is the NFW concentration parameter of the subhalo at
accretion, and $r_\rmh$ is the instantaneous half-mass radius. These two
criteria can be used to assess whether individual subhaloes in
cosmological simulations are reliable or not. In fact, we recommend
that subhaloes that satisfy either of these two criteria be discarded
from further analysis. 

As discussed at length in Paper~I, being able to accurately predict
the abundance and demographics of dark matter substructure is of
crucial importance for a wide range of astrophysics. First of all, it
is one of the prime discriminators between different dark matter
models; if dark matter is warm (WDM), rather than cold, free-streaming
and enhanced tidal disruption, will cause a significantly reduced
abundance of {\it low mass} subhaloes \citep[e.g.,][]{Knebe.etal.08,
  Lovell.etal.14, Colin.etal.15, Bose.etal.16}. If dark matter has a
significant cross-section for self-interaction (SIDM), subhaloes are
predicted to have constant-density cores, with significantly lower
central densities than their cold dark matter (CDM) counterparts
\citep[e.g.][]{Vogelsberger.etal.12, Rocha.etal.13}.  Dark matter
substructure also boosts the expected dark matter annihilation signal
\citep{Bergstrom.etal.99}, an effect that is typically quantified in
terms of a `boost-factor' \citep[e.g.,][]{Strigari.etal.07,
  Giocoli.etal.08b, Kuhlen.etal.08, Pieri.etal.08, Moline.etal.16}.
In addition, dark matter substructure is also important for
understanding galaxy formation and large scale structure.  Dark matter
subhaloes are believed to host satellite galaxies and their
demographics is therefore directly related to the (small scale)
clustering of galaxies. This idea underlies the popular technique of
subhalo abundance matching \citep[e.g.,][]{Vale.Ostriker.04,
  Conroy.etal.06, Guo.etal.10, Hearin.etal.13, Moster.etal.13,
  Behroozi.etal.13c}, which has become a prime tool for interpreting
galaxy clustering, galaxy-galaxy lensing, and group multiplicity
functions, and is even used to constrain cosmological parameters
\citep{Marin.etal.08, Trujillo-Gomez.etal.11, Hearin.etal.15,
  Hearin.etal.16, Reddick.etal.13, Reddick.etal.14, Zentner.etal.14,
  Zentner.etal.16, Lehmann.etal.15}.

The results presented here instill some serious concern regarding the
reliability of dark matter substructure in state-of-the-art
cosmological simulations. In particular, it questions whether the fact
that subhalo mass functions appear to be converged down to 50-100
particles per subhalo (see \S\ref{sec:intro}) implies that the results
are reliable. For example, as shown in Fig.~\ref{fig:r01}, even subhaloes
with as many as $10^6$ particles at accretion can experience
artificial disruption and/or discreteness noise. On the other hand,
although our results make it clear that the properties (e.g., mass,
size, maximum circular velocity, concentration) of {\it individual}
subhaloes can no longer be trusted once they violate
criterion~(\ref{disccrit}), it remains to be seen to what extent these
results impact {\it statistical} results, such as the subhalo mass
function. For example, subhaloes that violate
criterion~(\ref{disccrit}) have mass loss rates that are severely
affected by a discreteness noise driven runaway instability. In $\sim
50$ percent of the cases this results in the subhaloes loosing too
much mass, while the other half experience mass loss rates that are
too low. As a consequence, discreteness noise does not strongly affect
the {\it average} mass loss rate, and it remains to be seen whether
its overall effect on the subhalo mass (or velocity) function is
significant. We intend to address this issue in the near future (Ogiya et
al., in preparation) using a series of idealized, high-resolution,
converged simulations (similar to the converged simulations used here
to gauge the impact of numerics), that span a wide range of parameter
space (orbital energy and circularity, as well as the concentration 
parameters of the host and sub-halo). Such a data-base can be used to 
calibrate semi-analytical models describing the build-up and evolution 
of dark matter substructure \citep[e.g.,][]{Taylor.Babul.01, 
Penarrubia.Benson.05, vdBosch.etal.05, Zentner.etal.05, Jiang.vdBosch.16}, 
which in turn can be used to correct cosmological simulations for 
artificial subhalo disruption. This data-base will also be instrumental 
for informing and facilitating a more accurate modeling of `orphan 
galaxies' in semi-analytical models of galaxy formation
\citep[e.g.,][]{Springel.etal.01, Kang.etal.05, Kitzbichler.White.08},
subhalo abundance matching models \citep[e.g.,][]{Conroy.etal.06,
  Guo.White.13, Campbell.etal.17} and empirical models
\citep[e.g.,][]{Moster.etal.10, Moster.etal.17, Lu.etal.14}.  We refer 
the reader to \cite{Tollet.etal.17} and \cite{Pujol.etal.17} for detailed
discussions regarding the various methods that are being used to treat
such orphans.

We end with a brief discussion of how to proceed. In particular, how
can we improve numerical simulations such that overmerging is less of
an issue? The only way to overcome the discreteness driven runaway
instability is to increase the mass resolution (i.e., increase the
number of particles used in the simulations). As computer power
continues to increase, so will the mass resolution of the
state-of-the-art cosmological simulations. Consequently, at a fixed
mass, the resolution will undoubtedly continue to improve, thereby
suppressing issues related to discreteness noise. However, criterion~
(\ref{disccritINV}) will remain valid, and there will thus always be a
mass scale below which subhaloes are no longer treated
reliably. Overcoming the {\it systematic} issue of inadequate force
softening is far more challenging.  Ideally, the softening scale 
varies with both space and time. Few studies 
have attempted to implement such adaptive softening strategies in 
particle-based $N$-body codes. \cite{Price.Monaghan.07} and
\cite{Iannuzzi.Dolag.11} implemented schemes in which the softening
adapts to the local density (softening scale decreases with increasing
density), while retaining conservation of both momentum 
and energy. \cite{Hobbs.etal.16} even account for the anisotropy
of the local density field, in an attempt to overcome the problem
of artificial fragmentation that plagues simulations of structure
formation in warm dark matter cosmologies. It is unclear, though, 
whether any of these schemes will resolve the subhalo overmerging 
problem. As we have shown in \S\ref{sec:softening}, the optimal 
softening length rapidly declines once the subhalo starts to experience 
mass loss, and ideally subhaloes are therefore simulated with a softening 
length that is significantly {\it smaller} than that for host haloes 
of a similar mass. However, due to tidal shock heating and re-virialization, 
the central densities of subhaloes are typically smaller than those of 
their host haloes (see Fig.~\ref{fig:prof06}). In the adaptive softening 
schemes mentioned above, the central regions of subhaloes would therefore 
be resolved with a {\it larger} softening length than the central
region of a host halo. Until a
solution to overcome numerical overmerging is found, the only option
is to supplement the simulation results with a semi-analytical model.
In particular, using the criteria presented in this paper one first
identifies subhaloes the moment they become unreliable. Their
subsequent evolution is then treated using a semi-analytical model
that uses orbit integration, and treatments of tidal stripping, tidal
heating and dynamical friction, to predict their mass and
location. Unfortunately, as discussed at great length in Paper~I, we
lack a rigorous, analytical treatment for how the mass, structure, and
orbits of subhaloes evolve in the presence of a tidal field. These
semi-analytical models, therefore, need to be calibrated using
high-resolution, properly converged simulation results. We intend to
address this `catch-22' in the near future using the large date-base
of idealized simulations discussed above.


\section*{Acknowledgments}

This work has benefited greatly from discussions with the following 
individuals: Andy Burkert, Benedikt Diemer, Oliver Hahn, Andrew Hearin, 
Fangzhou Jiang, Johannes Lange, and Fred Rasio. We also acknowledge the
referee, Chris Power, for an insightful, constructive referee report.
FvdB is grateful to the Munich Excellence Cluster for its hospitality, 
and is supported by the Klaus Tschira Foundation and by the US
National Science Foundation through grant AST 1516962. Part of this
work was performed at the Aspen Center for Physics, which is supported
by the National Science Foundation under grant PHY-1066293, and at the
Kavli Institute for Theoretical Physics, which is supported by the
National Science Foundation under grant PHY-1125915.  GO is 
supported by funding form the European Research Council (ERC) under
the European Union's Horizon 2020 research and innovation programme
(grant agreement No. 679145, project `COSMO-SIMS'). 
Some of the numerical computations were carried out on HA-PACS at the 
Center for Computational Sciences at University of Tsukuba.


\bibliographystyle{mnras}
\bibliography{references_vdb}


\appendix

\section{Discreteness Noise in the Subhalo Mass Loss Rate}
\label{App:dis}

In \S\ref{sec:convergence} we discuss how discreteness noise gives
rise to a runaway instability, in which the variance in $\fbound(t)$
among many simulations, which only differ in their random realizations
of the initial conditions, increases with time. We defined $T_{\rm
  dis}$ as the time when the standard deviation in $\fbound$ among
these simulations becomes 0.1 dex, and defined $N_{\rm dis}$ as the
average number of bound particles in the subhalo at time $T_{\rm
  dis}$.  Here we derive a relation between $N_{\rm dis}$ and the
number of particles, $\Np$, in the initial subhalo, assuming only that
the discreteness noise obeys Poisson statistics.

Let $\langle \Delta N \rangle$ be the expectation value for the number
of particles that will be stripped off during a time interval $\Delta
t$. We have that $\langle \Delta N \rangle = \dot{m} \, \Delta
t/m_\rmp$, where $m_\rmp$ is the particle mass, and $\dot{m}$ is the
system's instantaneous mass loss rate in the limit $m_\rmp \rightarrow
0$.  Due to sampling noise, there will be Poisson fluctuations in the
actual $\Delta N$, and thus in the mass-loss rate in the numerical
simulation. In particular, the typical error in $\Delta N$ due to
these Poisson fluctuations will be equal to $\sigma_{\Delta N} =
\sqrt{\langle \Delta N \rangle}$.  Clearly, the only way to suppress
this Poisson noise is to increase the number of particles that is used
to model the subhalo.

Consider a system that, due to this Poisson noise, experiences a high
fluctuation in the mass loss rate during a time step $\Delta t$, i.e.,
$\Delta N > \langle \Delta N \rangle$. Subsequently, the system will
re-virialize (roughly on a dynamical time) to adjust to this mass
loss. Because the system in question experienced a higher-than-average
episode of mass loss, it will also experience more significant
re-virialization, which results in a more-than-average expansion of
the remnant. Consequently, the system in question will be less dense
than average, and hence experience more mass loss in the subsequent
time interval; after all, less dense systems experience higher mass
loss rates. Hence, the fluctuations in $\Delta N$ in different
time-intervals $\Delta t$ are going to be positively correlated, which
implies that the numerical error in the time-evolution of the subhalo
mass, $m(t)$, will grow with time; i.e., {\it discreteness noise in a
  tidal field results in a run-away instability}.

Here we use a simple toy model to estimate at what $N$ this problem
becomes appreciable. If we define `appreciable' as corresponding to a
situation in which a $1\sigma$ fluctuation in $\Delta N$ (during a
time-interval $\Delta t$) implies a fluctuation in the corresponding
mass-loss-rate of 10 percent, then we have that the problem becomes
appreciable if $\langle \Delta N \rangle = 100$. The relevant
time-scale $\Delta t$ is the time-scale on which the subhalo
re-arranges its mass distribution in response to the mass loss, which
is of order the dynamical time, $\tau_{\rm dyn}$, of the subhalo
remnant. Hence, we argue that the Poisson fluctuations become
appreciable once the expectation value for the mass loss rate becomes
of order 100 particles per subhalo dynamical time.

\cite{Jiang.vdBosch.16} have shown that the {\it average}
subhalo mass loss rate (averaged over all orbits and orbital phases)
in numerical simulations is such that, to good approximation,
\begin{equation}
m(t+\Delta t) = m(t)\,\exp(-\Delta t/\tau_{\rm strip})\,,
\end{equation}
where $\tau_{\rm strip}$ is a characteristic time scale for mass loss.
Using this relation, we find that a subhalo looses on average 100
particles per dynamical time if its instantaneous number of particles
is
\begin{equation}\label{Ndistau}
N = \frac{100}{1 - {\rm exp}[-\tau_{\rm dyn}/\tau_{\rm strip}]}\,.
\end{equation}

Using that the dynamical time is proportional to the crossing time,
\begin{equation}
  t_{\rm cross} \equiv \sqrt{\frac{2 \, r_\rmh^3}{G \, m(t)}}\,
\end{equation}
with $r_\rmh$ the subhalo's half-mass radius, and $m(t)$ the bound
mass of the subhalo at time $t$, we have that
\begin{equation}\label{tdyncross}
  \tau_{\rm dyn}(t) \propto t_{{\rm cross},0} \,
  \left[ \frac{(r_\rmh/r_{{\rm h},0})^3}{\fbound} \right]^{1/2}\,,
\end{equation}
where a subscript 0 refers to the initial property of the subhalo,
i.e., prior to being exposed to an external tidal field. In the case
of cosmological simulations, this roughly coincides with the
properties of the subhalo at the time of accretion into the host halo.

As discussed in \S\ref{sec:eps}, our simulations reveal a reasonably
tight relation between $\fbound$ and $r_\rmh/r_{{\rmh},0}$ of the form
$r_\rmh/r_{{\rmh},0} \propto \fbound^{0.5}$, independent of the
subhalo's orbit.  Substituting this in Eq.~\ref{tdyncross}, we can
rewrite Eq.~\ref{Ndistau} as
\begin{equation}\label{eqN}
N = \frac{100}{1 - {\rm exp}[-\alpha \fbound^{0.25}]}\,,
\end{equation}
where $\alpha \propto (\tau_{{\rm dyn},0}/\tau_{\rm strip})$ is a
unitless parameter. Taylor expanding Eq.~(\ref{eqN}) to first order
yields that $N \propto \fbound^{-0.25}$. If we assume that this
instantaneous number of particles is proportional to $N_{\rm dis}$,
then, using that $\fbound = N/\Np$, we finally obtain that $N_{\rm
  dis} \propto \Np^{0.2}$. As discussed in \S\ref{sec:Ndis}, this 
is exactly the behavior seen in our simulation data
(cf. Fig.~\ref{fig:crit}).

%
%
%
%
%
%
%

\label{lastpage}

\end{document}